\documentclass[]{aa} 

\usepackage{graphicx}
\usepackage{txfonts}
\usepackage{adjustbox}
\usepackage{multirow}
\usepackage{gensymb}

\usepackage{mwe}
\usepackage{subfig}
\usepackage{hyperref}
\usepackage{url}
\usepackage{float,capt-of}
\usepackage{natbib}
\usepackage[english]{babel}
\usepackage{graphics,rotating,amssymb,amsmath}
\usepackage{xcolor}
\usepackage[utf8]{inputenc}

\newcommand{\mr}{\multirow{2}{*}}
\newcommand{\mc}{\multicolumn{2}{c}}
\newcommand{\fmp}{\texttt{FMP}}
\newcommand{\Ha}{H$\alpha$}
\newcommand{\Hb}{H$\beta$}
\newcommand{\Hc}{H$\gamma$}

\begin{document}

\title{GMP-selected dual and lensed AGNs: selection function and classification based on near-IR colors and resolved spectra from VLT/ERIS, KECK/OSIRIS, and LBT/LUCI}

\titlerunning{Classification of GMP systems}
\authorrunning{Mannucci et al.}


\author{
F. Mannucci \inst{1}\thanks{\email{filippo.mannucci@inaf.it}}
\and M. Scialpi \inst{4,1}
\and A. Ciurlo\inst{2}
\and S. Yeh\inst{3}
\and C. Marconcini \inst{4,1}
\and G. Tozzi\inst{4,1}
\and G. Cresci\inst{1}
\and A. Marconi\inst{4,1}
\and A. Amiri\inst{5,4}
\and F. Belfiore \inst{1}
\and S. Carniani \inst{6}
\and C. Cicone \inst{7}
\and E. Nardini \inst{1}
\and E. Pancino \inst{1}
\and K. Rubinur\inst{7}     
\and P. Severgnini \inst{8}
\and L. Ulivi \inst{4,1}
\and G. Venturi \inst{6,1}
\and C. Vignali \inst{9}
\and M. Volonteri \inst{10}
\and E. Pinna \inst{1}
\and F. Rossi \inst{1}
\and A. Puglisi \inst{1}
\and G. Agapito \inst{1}
\and C. Plantet \inst{1}
\and E. Ghose \inst{1}
\and L. Carbonaro \inst{1}
\and M. Xompero \inst{1}
\and P. Grani \inst{1}
\and S. Esposito \inst{1}
\and J. Power \inst{11}
\and J. C. Guerra Ramon \inst{11}
\and M. Lefebvre \inst{11}
\and A. Cavallaro \inst{11}
\and R. Davies \inst{12}
\and A. Riccardi \inst{1}
\and M. Macintosh	\inst{13}
\and W. Taylor	\inst{13}
\and M. Dolci	\inst{14}
\and A. Baruffolo	\inst{15}
\and H. Feuchtgruber\inst{12}
\and K. Kravchenko	\inst{12}
\and C. Rau	\inst{12}
\and E. Sturm	\inst{12}
\and E. Wiezorrek	\inst{12}
\and Y. Dallilar	\inst{12,16}	
\and M. Kenworthy	\inst{17}
}

\institute{
INAF - Osservatorio Astrofisico di Arcetri, largo E. Fermi 5, 50125 Firenze, Italy.  \and 
Department of Physics and Astronomy, University of California Los Angeles, 430 Portola Plaza, Los Angeles, CA 90095, USA.  \and 
W. M Keck Observatory, 65-1120 Mamalahoa Highway, Kamuela, HI 96743, USA. \and 
Dipartimento di Fisica e Astronomia, Università di Firenze, Via G. Sansone 1, 50019, Sesto Fiorentino (Firenze), Italy. \and 
Department of Physics, University of Arkansas, 226 Physics Building, 825 West Dickson Street, Fayetteville, AR 72701, USA. \and
Scuola Normale Superiore, Piazza dei Cavalieri 7, 56126, Pisa, Italy. \and 
Institute of Theoretical Astrophysics, University of Oslo, P.O Box 1029, Blindern, 0315 Oslo, Norway. \and 
INAF - Osservatorio Astronomico di Brera, Via Brera 28,20121 Milano, Italy \and 
Physics and Astronomy Department "Augusto Righi", Università di Bologna, Via Gobetti 93/2, 40129 Bologna, Italy. \and 
Institu d'Astrophysique de Paris,  98bis Bd Arago, 75014 Paris, France. \and
Large Binocular Telescope Observatory, Tucson, Arizona, USA 
\and
Max-Planck-Institut f\:ur extraterrestrische Physik, Postfach 1312, 85741, Garching, Germany
\and
STFC UK ATC, Royal Observatory Edinburgh, Blackford Hill. Edinburgh, EH9 3HJ, UK 
\and
INAF – Osservatorio Astronomico d’Abruzzo, Via Mentore Maggini, 64100, Teramo, Italy
\and
INAF – Osservatorio Astronomico di Padova, Vicolo dell’Osservatorio 5, 35122, Padova, Italy
\and
I. Physikalisches Institut, Universität zu Köln, Zülpicher Str. 77, 50937, Köln, Germany
\and
Leiden Observatory, University of Leiden, P.O. Box 9513, 2300 RA Leiden, The Netherlands
}


\date{}

\abstract
{
The Gaia-Multi-Peak (GMP) technique can be used to identify large numbers of dual or lensed active galactic nuclei (AGN) candidates at sub-arcsec separation, allowing us to study both multiple super-massive black holes (SMBHs) in the same galaxy and rare, compact lensed systems. The observed samples can be used to test the predictions of the models of SMBH merging once 1) the selection function of the GMP technique is known, and 2) each system has been classified as dual AGN, lensed AGN, or AGN/star alignment. Here we 
show that the GMP selection is very efficient for separations above $0.15\arcsec$ when the secondary (fainter) object has magnitude G$\lesssim$20.5.
We present the spectroscopic classification of five GMP candidates using VLT/ERIS and Keck/OSIRIS, and compare them with the classifications obtained from: a) the near-IR colors of 7 systems obtained with LBT/LUCI, and b) the analysis of the total, spatially-unresolved spectra. We conclude that colors and integrated spectra can already provide reliable classifications of many systems. Finally, we summarize the confirmed dual AGNs at z$>$0.5 selected by the GMP technique, and compare this sample with other such systems from the literature, concluding that GMP can provide a large number of confirmed dual AGNs at separations below 7~kpc. 
}

\keywords{
           }

\maketitle
%

\section{Introduction}

The existence of dual AGNs,  i.e., pairs of active supermassive black holes (SMBH) at kpc separations inspiralling in the same galaxy, is one of the fundamental predictions of current models of galaxy and SMBH formation and evolution \citep[e.g.][]{Tremmel17}. 
The study of these systems allows us to test the models of formation and evolution of binary SMBHs. These systems are particular important because 
mergers of SMBHs in the mass range $10^4-10^9~\rm{M}_\odot$ will dominate the gravitational wave (GW) events detected by the future Laser Interferometer Space Antenna (LISA) space mission and by the Pulsar Timing Arrays \citep[e.g.,][]{Arzoumanian18,Amaro-Seoane23}. 
Dual AGNs allow us to investigate the starting conditions of the merging process, and interpret future GW results \citep[e.g.][]{DeGraf23,Dong-Paez23,Chen22c}.

Models of galaxy and SMBH co-evolution predict that dual AGNs at small separations (less than 10~kpc) are quite common, 
reaching a few percent of all existing AGNs at $z>0.5$, where merging activity is expected to be larger than in the local Universe \citep{Volonteri21} and therefore must exist in a significant fraction of galaxies. Despite these expectations and a decade of active searches, only a handful of confirmed dual AGNs at small separations are currently known  \citep[see][for a recent compilation of the existing data]{Chen22b}.

These objects can only be detected through imaging with a spatial resolution considerably better than one arcsec, typically requiring observation from space \citep[see][for review]{derosa20}. Being rare systems, they also require very wide field surveys. 
The ESA mission Gaia 
\citep{Prusti16} is currently the only mission providing high-resolution data over large areas of the sky, therefore can be exploited to detect dual and lensed AGNs. 
The presence of multiple Gaia sources associated, within a few arcsec, to known AGNs
(the "multiplicity" method) have been used 
by several authors
\citep[e.g.][]{Lemon18,Chen22a}
to detect both dual and lensed systems.
The extra astrometric jitter induced by AGN variation in unresolved pairs was also exploited to select several multiple AGN candidates
\citep[e.g.][]{Shen19,Hwang20,Chen22a}.

Recently, in \cite{Mannucci22},  we developed the GMP method which exploits the Gaia catalog to select a large number of dual AGN candidates at sub-arcsec separations. This method is based on the detection of multiple peaks in the light profiles of otherwise unresolved AGNs. 
Using spatially-resolved imaging and spectroscopy, we showed that the GMP is a very effective method to select dual systems, while in  \cite{Ciurlo23} and \cite{Scialpi23} we built the first sample of 
spectroscopically confirmed, GMP-selected, dual AGNs at sub-arcsec separations.

The GMP technique is expected to provide a large number of confirmed dual AGNs, especially at $z>0.3$ where the host galaxy does not contribute significantly to the Gaia light profile. These results will be compared with the expectations of the models of galaxy/SMBH formation and co-evolution, which predict various quantities such as the distribution in separation, luminosity and mass ratio, redshift, etc \citep{Steinborn2016,Capelo2017,Rosas-Guevara2019,Volonteri21,Chen22c}. The samples of GMP-selected dual AGNs will provide strong constraints on these models once the efficiency of this method in selecting multiple systems is known as a function of separation and luminosity of the components.\\

QSOs showing multiple components need to be classified as either a pair of AGNs (either a dual system or a single AGN lensed by a foreground galaxy) or an alignment between an AGN and a foreground star. Different possibilities exist:
\begin{enumerate}
\item The classification is usually obtained by spatially resolved spectroscopy from  space \citep[e.g][]{Junkkarinen01, Mannucci22} or by adaptive-optics (AO) assisted spectroscopy from the ground \citep[e.g][]{Ciurlo23, Scialpi23}. The result classification can be very reliable, but obtaining resolved spectroscopy is very telescope-expensive and limits the number of confirmed systems. 
\item Spatially unresolved spectroscopy, usually from the ground in seeing-limited conditions, can be analyzed to detect stellar features at zero velocity, revealing the presence of a star projected close to the AGN \citep{Shen23,Scialpi23}. These data cannot distinguish dual and lensed systems, since in both cases the spectral differences are not expected to be large enough to be seen in the total spectrum. 
\item Since stars and AGNs have, in most cases, different colors, multi-band spatially-resolved imaging can be used to study the nature of the two components \citep[e.g.][]{Chen22a,Chen22b}. Intrinsic luminosities can also be used to identify lensed systems, in many cases much brighter than the typical AGN at the same redshift.
\end{enumerate}

By applying these three different classification methods to the same targets, in this paper we show that spatially-resolved imaging and spatially-unresolved spectroscopy can complement resolved spectroscopy in classifying at least part of the systems, distinguishing between dual AGNs, lensed systems, and star/AGN alignments.

In Sect.~\ref{sec:effic} we discuss the statistical properties of the GMP-selected systems in terms of luminosity ratios and projected separation. 
In Sect.~\ref{sec:classif} we present and compare the different classification techniques.  
In Sect.~\ref{sec:colors} we describe the Large Binocular Telescope (LBT) observations of seven targets aimed at measuring the near-IR colors of each component.
Sect.~\ref{sec:deconv} describes how the analysis of the integrated spectra can reveal the presence of stars projected close to the AGN.
In. Sect.~\ref{sec:AOspectra} we present spatially-resolved ESO Very Large Telescope (VLT) and Keck spectra of five GMP-selected targets, which include two confirmed dual AGNs and one compact lensed system.
In Sect.~\ref{sec:comparison} we compare the results of the three classification.
Finally, in Sect.~\ref{sec:summary} we summarize the state of GMP target observations and discuss the results in the broader context of observed dual AGNs at small separations.


\begin{figure*}
\centering
\includegraphics[width=0.47\textwidth]{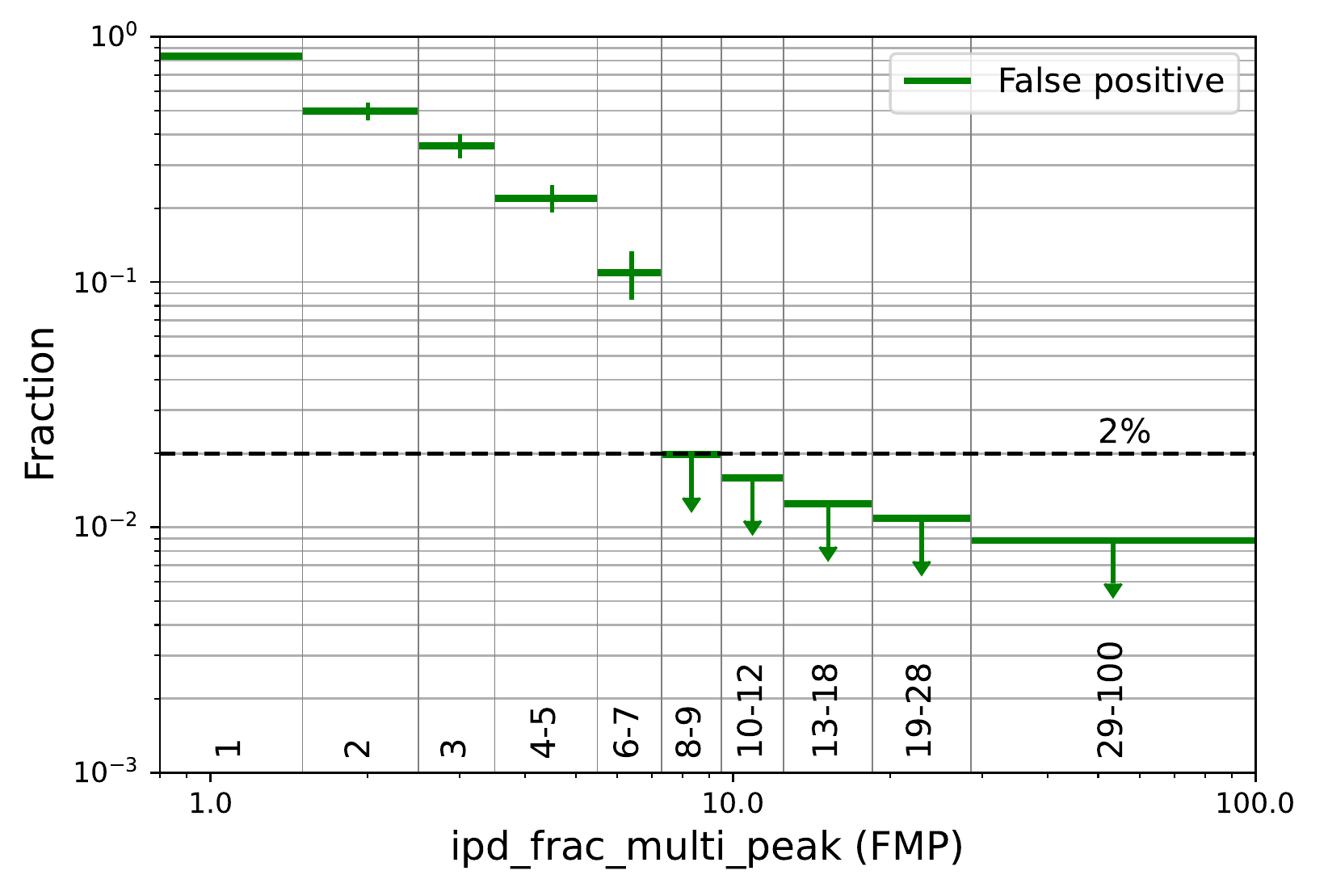}
\includegraphics[width=0.49\textwidth]{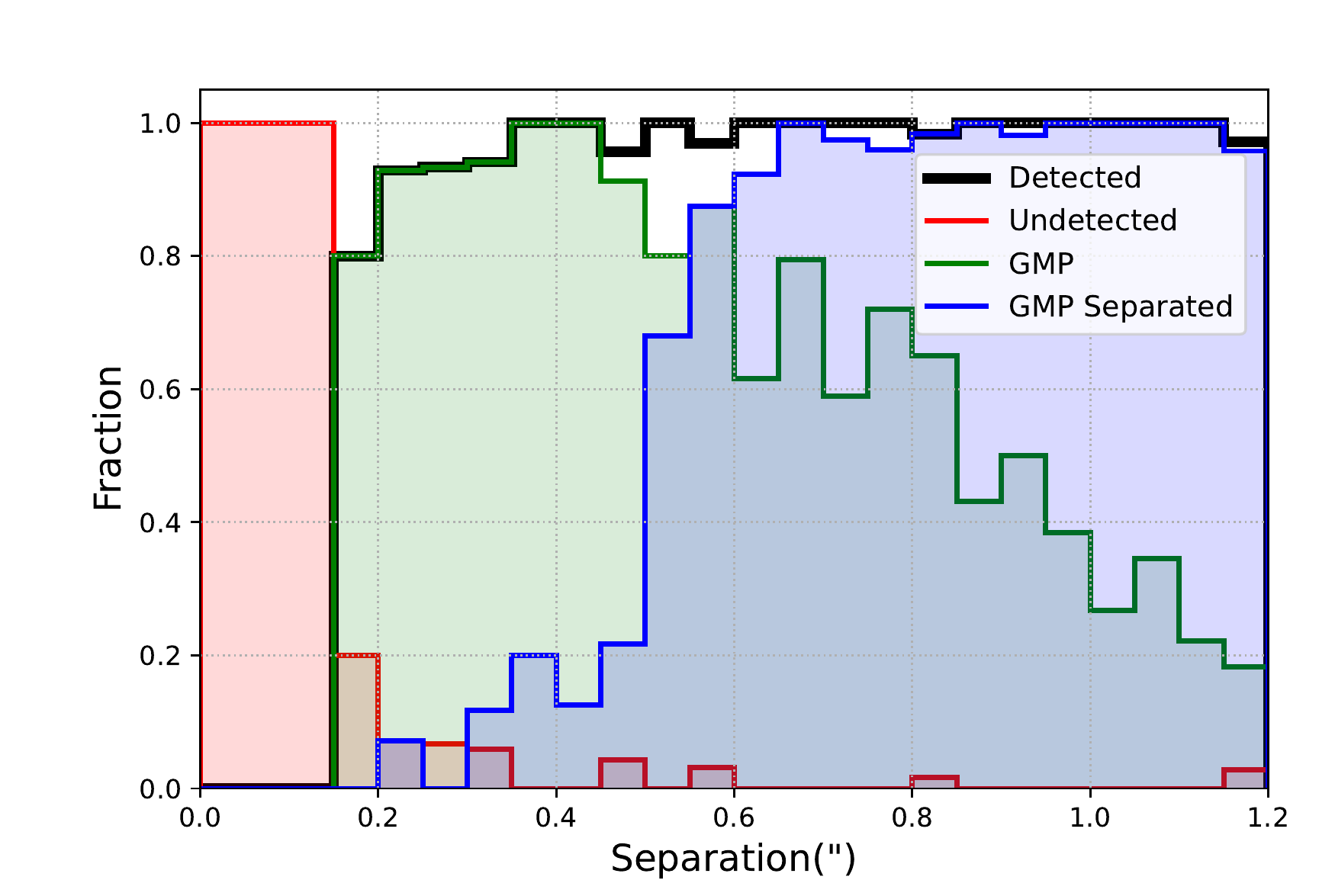}

\caption{
{\em Left:} Level of contamination of false positives (objects with detection of multi peaks that are not projected pairs) as a function of the \fmp\ values in a sample of dense stellar fields, for G$<20.5$. No false positives are found for \fmp$\geq8$, where we plot 90\% upper limits, i.e., the levels that would correspond to 2.3 false positive.
{\em Right:} fraction of recovered projected pairs as a function of separation ($\delta$) in arcsec. The red histogram shows the pairs that are not recovered, which dominate the counts at $\delta<0.15\arcsec$ and are very rare at $\delta>0.20\arcsec$. The green histogram reports the fraction of recovered `isolated' objects that correspond to single entries in the Gaia source catalog, and the blue one shows the GMP-selected objects that have nearby companions in the catalog. The combination of these two GMP populations is show by the black line.
}

\label{fig:effic}
\end{figure*}

\begin{figure*}
\centering
\includegraphics[width=0.95\textwidth]{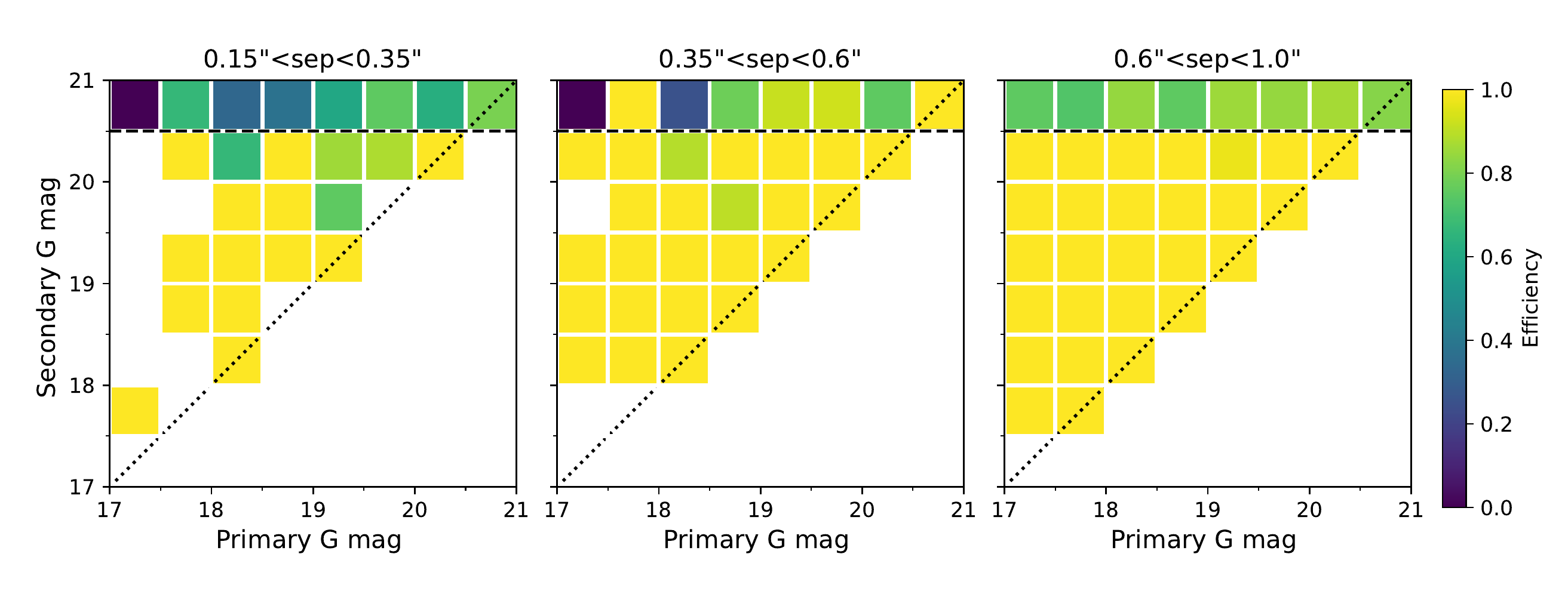}

\caption{Efficiency of the GMP method as a function of
G mag of the primary and 
primary and secondary components, in three separation ranges. The efficiency does not show strong dependencies on the three parameters up to secondary G magnitudes of 20.5, shown by the dashed line.
}

\label{fig:effic2D}%
\end{figure*}

\section{GMP selection efficiency and contamination}
\label{sec:effic}

The GMP technique is based on the detection of multiple peaks in the light profiles
of known AGNs in the multiple Gaia scans. The Gaia DR3 catalog \citep{Vallenari22} provides the parameter $ipd\_frac\_multi\_peak$\ (\fmp) that contains the percent fraction of scans showing a multiple peak for a given target. 
The value of this parameter, which ranges from 0 (multiple peak never detected) to 100 (always detected), is determined by several factors. Separation and luminosity ratios are important parameters to consider, as pairs with too small separations and/or too large luminosity ratios will not produce a measurable double peak. Since Gaia obtains sky scans in multiple directions, an object pair will not show a double peak when the scanning direction is perpendicular to the separation line. In addition, when scanning objects with separations  $\delta$ larger or of the order of the Gaia photometric window  (0.7$\arcsec$, \citealt{Prusti16}), the components may be separated into different catalog entries, each one with a single peak and a low value of \fmp. In contrast, since Gaia projects on the same array two different fields $106.5\degree$ apart \citep{Prusti16},
double peaks can be occasionally detected in light profile of isolated objects due to the alignment with a source in the other field. Thus, the selection requires a statistical approach, and needs to tackle the following complementary challenges:
\begin{enumerate}
\item Define the minimum value of \fmp\ to keep the level of {\it contamination} by false positives (single systems with values of \fmp\ above threshold) low. 
\item Measure the {\it efficiency} of the GMP method
of detecting multiple objects as a function of magnitudes and angular separations $\delta$ of the components. 
\end{enumerate}
These two problems can be addressed by testing what values of \fmp\ are obtained by apparent stellar pairs in moderately crowded stellar fields as a function of their magnitude and separations. For this, we have used the 
stellar catalogs\footnote{\href{http://groups.dfa.unipd.it/ESPG/hstphot.html}{http://groups.dfa.unipd.it/ESPG/hstphot.html}} 
of 110 parallel fields observed with HST in the F475W and F814W filters in the outskirts of 48 globular clusters (GCs) by \cite{Piotto15} and \cite{Simioni18}. These images have depth and spatial resolution better than Gaia and, as a consequence, can be used to obtain a sample of projected pairs and test the performance of the GMP method. 
Even if the central parts of the GCs are avoided, some of these fields are very crowded. We have limited our analysis to the 84 fields having densities such that the average distance of a source to the closest object is above $5\arcsec$.
Higher densities tend to have too many objects at close projected distances, a regime that is not present in the high galactic latitude fields used for the GMP selection, and that could affect Gaia detection properties. Restricting to lower densities would result in a low number of projected pairs at low separation. Regions within $5\arcsec$ of bright stars (G$<$15.5) were excluded because the rectangular Gaia primary mirrors produce long diffraction spikes that could produce the detection of secondary peaks in nearby sources. 

\subsection{Contamination}
The level of contamination by false positives, i.e. objects with high values of \fmp\ that are not projected pairs at the HST resolution ($\sim0.05\arcsec$), can be estimated by considering all the objects in the observed fields that have \fmp\ above a given threshold in the Gaia catalog, and by investigating what fraction does not correspond to pairs in the HST catalog. The results are shown in the left panel of Fig.~\ref{fig:effic}. While at \fmp=1 the fraction of false positives is about 80\%, no false positives are found with \fmp$\geq8$ far from bright stars.
The maximum fraction of false positives is at the percent level in each of the bins with \fmp$\geq8$ shown in Fig.~\ref{fig:effic}, and is below $\sim2\times10^{-3}$ considering all of them together.

\begin{table*}[t]
\centering
\begin{tabular}{ccccccccc}
\hline
\hline
 Target     &   RA        &    DEC      &  ~$z$ &\fmp&   G         &   Integ.  &  Resolved  & Resolved\\
            &             &             &       &    &             & spectrum  &  colors   & spectra \\
\hline
J0732+3533  & 07:32:51.57 & +35:33:15.3 & 3.065 & 10 & 20.31       & DR16       & LBT/LUCI  &   -    \\
J0812--0040 & 08:12:19.34 &--00:40:47.9 & 1.912 & 20 & 20.36       & DR16       & LBT/LUCI  &   -    \\
J0812+2007  & 08:12:46.41 & +20:07:30.1 & 1.48  & 23 & 20.16/20.44 & DR16       & LBT/LUCI  &   Keck/OSIRIS \\
J0927+3512  & 09:27:48.42 & +35:12:41.3 & 1.149 & 49 & 15.86       & LAMOST     & LBT/LUCI  &   -      \\
J0950+4329  & 09:50:31.63 & +43:29:08.6 & 1.770 & 78 & 17.90       & DR16       & LBT/LUCI  &   Keck/OSIRIS \\
J1048+4541	& 10:48:20.91 & +45:41:41.3 & 1.441 & 13 & 20.02       & DR16       &   -       &   Keck/OSIRIS \\
J1103+2348	& 11:03:08.10 & +23:48:05.8 & 1.441 & 17 & 20.16       & DR16       &   -       &   Keck/OSIRIS \\
J1318--0136 & 13:18:58.73 &--01:36:42.5 & 1.486 & 17 & 20.03       & 2QZ        & LBT/LUCI  &   VLT/ERIS  \\
J1510+5959  & 15:10:45.51 & +59:59:19.0 & 2.004 & 34 & 20.34       & DR16       & LBT/LUCI  &   -     \\
\hline
\end{tabular}
\caption[]{Main properties of the targets used to compare the various techniques, and source of the observational data. For each object we report the total Gaia G-band magnitude, with the exception for J0812+2007, which is separated into two targets in the Gaia catalog, and or which we report the magnitude of both objects. Integrated spectra are from SDSS DR16 \citep{Lyke20}, LAMOST \citep{Yao19}, and 2QZ \citep{Croom04}.
}
\label{tab:summary}
\end{table*}

\subsection{Efficiency}
The efficiency of pair detection can be estimated by the ratio of the number of recovered pairs vs. the total number of pairs present in the HST data, as a function of magnitude and separation $\delta$. Our catalog contains about 1300 projected pairs with $\delta<1.2\arcsec$ and with primary G$<$21~mag. The fraction of recovered objects (\fmp$>$8) with G$<$20.5~mag as a function of separation is shown in the right panel of Fig.~\ref{fig:effic}. The pairs that are not recovered (red histogram) dominate the counts at $\delta<0.15\arcsec$. Above this limit, virtually all the projected pairs are recovered within our magnitude limit. GMP objects that are isolated in the Gaia catalog, i.e., with no other object within $1.5\arcsec$ of separation, are shown in green. System corresponding to two nearby entries in the catalog, and that therefore can also be identified without the use of the GMP method, are shown in blue. The "isolated" population dominates separations between $0.15\arcsec$ and $0.6\arcsec$, while above this limit most of the objects are separated in the catalog. Of course, the "separated" population can also be identified by simply looking for pairs of objects at close separations in the Gaia archive, without considering the \fmp\ parameter. In particular, this population becomes dominant at $\delta>0.6\arcsec$, while below this limit most of the targets are "isolated" and can only be recovered by the GMP method.

For G$<$20.5~mag and $\delta>0.15\arcsec$, the selection efficiency does not depend critically on magnitude and separation. This is shown in Fig.~\ref{fig:effic2D}, where we plot the efficiency as a function of both primary and secondary magnitudes at different separations. The estimate of the G-band magnitude of the secondary stars undetected by Gaia is derived from the  photometry in the HST F814W band by applying a color correction derived from the detected sources. Provided that the fainter object has G$<$20.5~mag and the separation is $\delta>0.15\arcsec$, the efficiency is high and not strongly dependent on these parameters. This is a critical point that will allow us to compare the observational results with the models. 
Fig.~\ref{fig:effic2D} also shows that, if the primary component has G$>$17~mag,  the GMP method can efficiently select systems with a secondary as faint as G=20.5~mag. The range of brighter magnitudes, G$<$17~mag, is not sampled in our fields but it is not very relevant because the majority of AGNs at $z>0.5$ are fainter than this limit. As a consequence, the range of luminosity ratio sampled by the GMP method depends on the luminosity of the primary source.
For G$=$18.5~mag, a typical value for bright AGN at $z>1$, the magnitude difference with the secondary member is limited to $\sim$2~mag, a factor of $\sim 6$ in luminosity. This limits the number of detected sources but guarantees that both components give some contribution to the integrated spectra, as discussed in the next section. Larger luminosity ratios can be covered with the ESA satellite EUCLID 
\citep{Scaramella22}. Multiple AGN detections with Euclid will be discussed in a future paper
(Ulivi et al., in preparation).


\section{Classifying multiple systems in different ways}
\label{sec:classif}

Since the GMP technique only provides the evidence of the presence of multiple components in an AGN, the classification of these components requires additional observations. As shown in \cite{Mannucci22}, \cite{Ciurlo23}, and \cite{Scialpi23}, resolved spectroscopy is the ultimate technique but is observationally very demanding and requires high spatial resolution. For this reason, in this section we investigate how spatially-resolved colors (Sect.~\ref{sec:colors}) and total spectra (Sect.~\ref{sec:deconv}) can provide a classification of the components. 
To do this, we have
considered a sample of nine GMP-selected systems (see \ref{tab:summary}) with archival total spectra. 
We have obtained spatially-resolved colors of seven of them with LBT/LUCI, and 
spatially resolved spectra of five of these systems with Keck/OSIRIS and VLT/ERIS. Target selection is described in Sect.\ref{sec:lbt} and \ref{sec:AOspectra} for the LBT and Keck/VLT samples, respectively. The classifications based on these three data sets are compared in Sect~\ref{sec:comparison}.

\begin{table*}[]
\centering
\begin{tabular}{cccccc}
\hline
\hline
 Target     &Sep($\arcsec$)& \multicolumn{2}{c}{Primary} & \multicolumn{2}{c}{Secondary}\\
            &              & $H$ mag   & $K_s$ mag   & $H$ mag   & $K_s$ mag \\ 
\hline
J0732+3533  & 0.52 & 18.02$\pm$0.05&  17.69$\pm$0.05 & 18.63$\pm$0.05&  18.13$\pm$0.05  \\
J0812--0040 & 0.66 & 19.58$\pm$0.05&  19.58$\pm$0.05 & 20.32$\pm$0.05&  19.64$\pm$0.05 \\
J0812+2007  & 0.54 & 19.57$\pm$0.30& 18.56$\pm$0.05 & 19.63$\pm$0.30&  19.21$\pm$0.05 \\
J0927+3512  & 0.33 & 14.40$\pm$0.05& 14.42$\pm$0.05 & 15.38$\pm$0.05& 15.30$\pm$0.05 \\
J0950+4329  & 0.37 & 14.98$\pm$0.05& 14.34$\pm$0.05 & 15.47$\pm$0.05&  14.98$\pm$0.05  \\
J1318--0136 & 0.27 & 18.83$\pm$0.20& 18.62$\pm$0.10 & 18.94$\pm$0.20&  18.61$\pm$0.10 \\
J1510+5959  & 0.28 & 19.57$\pm$0.15& 19.27$\pm$0.10 & 20.20$\pm$0.15& 19.46$\pm$0.10 \\
\hline
\end{tabular}
\caption[]{Main properties of the targets observed with LBT.  
}
\label{tab:LBTlist}
\end{table*}
%

\subsection{Classification using near-IR colors}
\label{sec:colors}

\subsubsection{LBT imaging}
\label{sec:lbt}
We obtained Adaptive-optics (AO) assisted imaging of seven GMP-selected systems at the LBT, including the five multiple systems presented in \cite{Mannucci22}.
We extracted the targets from the Milliquas catalog v7.2 \citep{Flesch21}, selecting objects with secure AGN classification and redshifts, and within $\sim30\arcsec$ of suitable natural-guide stars (NGS) to drive the AO system. After cross-correlating with the Gaia archive, we selected systems with \fmp$\geq10$, a more conservative limit that the value \fmp$\geq8$ defined in Sect.~\ref{sec:effic}. Six out of seven targets are 'isolated', i.e., have no companion in the Gaia catalog, while J0812+2007 has a known companion 0.66$\arcsec$ apart.

Observations were performed on March 9th 2022 for objects \object{J0732+3533}, \object{J0812$-$0040}, \object{J0812+2007}, \object{J0927+3512}, and \object{J0950+432}, and on June 3rd 2022 for \object{J1318$-$0136} and \object{J1510+5959}. 
We used the 
LUCI1 imager and spectrograph \citep{Mandel07} with the SOUL AO module \citep{Pinna16, Pinna21}. 
Each system was observed in the near-IR $H$ and $K_s$ filters with integration times of 
6~min in $H$ and 24~min in $K_s$ in the March run, and 15~min each in $H$ and $K_s$ for the June run.  In all cases, we used an image sampling of 15~mas~pixel$^{-1}$. Data reduction was performed with our custom made python reduction routine (PySNAP). The seeing was rather variable during the June 2022 run, and we automatically rejected single images showing a
point-spread-function (PSF) significantly worse than the average. 

All the systems show two components at sub-arcsec separation. The $K_s$-band imaging of these sources was presented in \cite{Mannucci22}. In Table~\ref{tab:LBTlist} we summarize the observed magnitudes and the main properties of the observed objects. All the sources are well resolved in both $H$ and $K_s$ bands, except for J0812+2007 in the $H$-band (see below). This image was obtained under bad seeing conditions, and the resulting PSF (FWHM$\sim0.65\arcsec$) is actually larger than the separation ($0.54\arcsec$). As a consequence, the relative photometry of the two components (see below) in this band is unreliable.

Photometry was performed by PSF fitting. Since the two sources in each system are at very small distance, they are expected to have the same PSF, thus we fitted two elliptical Gaussian functions with the same shape (FWHM and position angle), centers defined on the image with the best resolution, and free normalization. This procedure provides reliable values of the parameters even when the PSF FWHM is not significantly smaller than the separation, and the sources are not well resolved. 
Since two systems are bright enough to be in the 2MASS catalog \citep{Skrutskie06}, we used their total magnitudes to obtain an absolute flux calibration. 
In all cases where the sources are well resolved, the uncertainties are dominated by flux calibrations, estimated to be $\Delta({\rm mag})\sim$0.05.  J1318-0136, and J1510+5959 are the two most compact systems, with separations of 0.27$\arcsec$ and 0.28$\arcsec$, similar to the PSF FWHM. In these cases we estimate larger photometric errors of 0.10 and 0.15 mag. For the bad-seeing $H$-band photometry of  J0812+2007 we estimate errors on the relative photometry of $\Delta({\rm mag})\sim$0.30.

\begin{figure*}[t]
\centering
\includegraphics[width=0.9\textwidth]{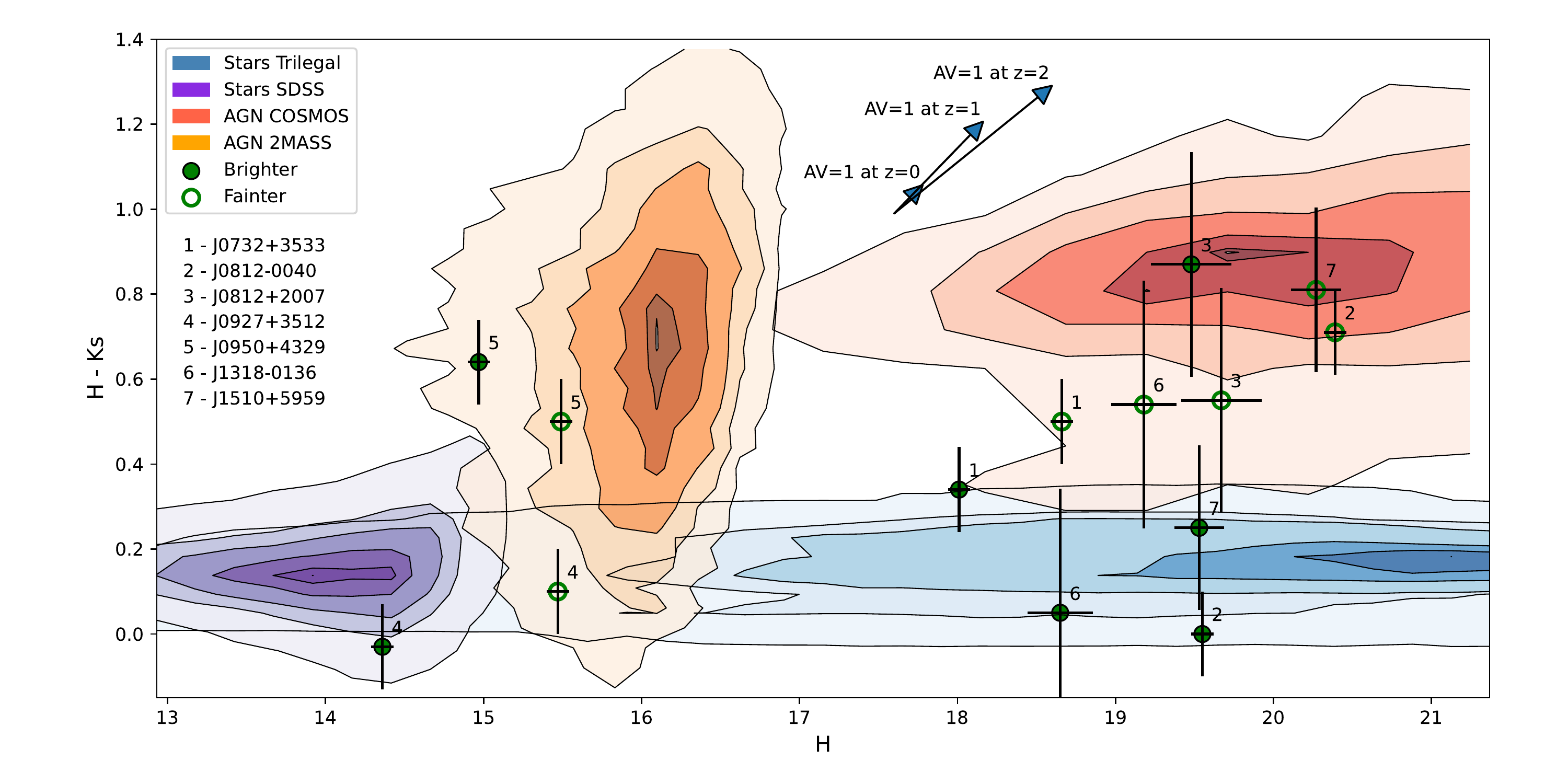}
\caption{Near-IR colors and magnitudes of the observed targets. Blue and purple contour plots show the distribution of Galactic stars from the TRILEGAL model \citep{Girardi05} and from the SDSS, respectively, while orange and red contours are AGNs from SDSS DR16Q \citep{Lyke20}
and COSMOS \citep{Civano16}, respectively.  
}
\label{fig:NIRcolors}%
\end{figure*}

\subsubsection{Spatially resolved colors}
\label{sec:IRcolors}

Fig.~\ref{fig:NIRcolors} shows the color-magnitude diagram for each component of the systems of the LBT sample, compared with QSOs and stars from the literature. For QSOs we used the 2MASS magnitudes of the 
SDSS DR16Q quasar catalog\footnote{\href{https://www.sdss.org/dr16/algorithms/qso\_catalog/}{https://www.sdss.org/dr16/algorithms/qso\_catalog/}} \citep{Lyke20}, 
and the much fainter QSOs in COSMOS by 
\cite{Civano16}\footnote{\href{https://cdsarc.cds.unistra.fr/viz-bin/cat/J/ApJ/819/62}{https://cdsarc.cds.unistra.fr/viz-bin/cat/J/ApJ/819/62}}. 
These two catalogs probe the entire luminosity range covered by the current sample ($15<H<21$). For stars we used observed magnitudes of a high galactic latitude star sample from the SDSS, and the model magnitudes of high galactic latitudes stars from the TRILEGAL Galaxy model \citep{Girardi05,Vanhollebeke09}. Stars show relatively blue $H-K_s$ colors of about 0.0--0.2, while QSOs span a larger range and mostly have redder colors $0.4<H-K_s<1.2$ with little evolution with magnitude (and therefore redshift).

The observed colors may be affected by the presence of dust in the AGN or in its host galaxy, whose effect  depends not only on its column density, but also on the redshift of the QSO, since our images sample progressively bluer wavelengths at higher redshift. As a reference, Fig.~\ref{fig:NIRcolors} shows the extinction arrows for $A_V=1$ for three different values of redshift.


The 14 components of the 7 LBT systems span a large part of this diagram.
Information on the nature of each component can be derived from its position in this color-magnitude diagram, i.e, by comparing it with the distribution of stars and QSO. A very small fraction of high-galactic latitude stars have colors $H-Ks\gtrsim0.35$, therefore objects with significantly redder colors can be identified as QSO, while object with bluer colors are most likely to be stars.
For all systems, except for \object{J0812+2007}, we do not know which one dominates the optical spectrum and is thus responsible for the classification as a QSO.

One component of J0812--0040, J1318-0136, and J1510+5959 (systems 2, 6, and 7 in Fig.~\ref{fig:NIRcolors}) show combinations of very blue colors and $H$-band magnitudes that are consistent with stars, while the other component has red $H-K_s$ colors typical of AGNs. These systems are therefore better described by AGN/star alignments, occurrences expected to be present in our sample at the 30\% level for separations larger than $0.5\arcsec$ \citep{Mannucci22}. It is very unlikely, although not totally impossible, that these systems are constituted by two AGNs.

J0732+3533 and J0812+2007 (systems 1 and 3) have one component compatible with being an AGN, and the other component consistent with both AGNs and stars. As a consequence, colors cannot provide a classification of these two objects.

Both components of J0950+4329 (system 5)  have colors that are only compatible with QSOs, and therefore this is either a dual or a lensed AGN.  This is a very bright object at $z=1.77$, at the upper luminosity envelope of the bright QSOs at this distance in \cite{Croom09}, and several magnitudes brighter than the typical COSMOS QSOs at that redshift \citep{Civano16}. For this reason, J0950+4329 is most likely a lensed AGN, whose luminosity is boosted by lensing magnification. The slightly different observed $H-K_s$ colors of the two components could be due to different amounts of extinction in the lensing galaxy. Given the separation and the redshift of the system, the lensing galaxy is expected to have $H\sim19-20$ and therefore would be outshined by the lensed QSOs.

Finally, both components of J0927+3512 (system 4) show very blue $H-K_s$ colors; the brighter component is only compatible with being a star, while fainter one could be either a star or a very bright (for its tabulated redshift of 1.149) and very blue AGN. Magnitude and colors of this system are more compatible with being a double star, despite its classification as an AGN based on the LAMOST spectrum. This point is further discussed in Sec.~\ref{sec:deconv}.

Objects fully resolved by Gaia, i.e., having a value of G magnitude for each component, can also be classified in a color-color plane involving both optical and near-IR magnitudes. This would offer the advantage of sampling a larger wavelength range and would provide a better classification of any companion star. In our sample, only 
\object{J0812+2007} is among these resolved targets, but the large error of its $H$-band magnitude prevents us from deriving any additional useful information.

In Sect.~\ref{sec:comparison} these classification based on resolved colors will be compared with those obtained from both total and spatially resolved spectroscopy derived in the next sections.

\begin{figure*}
\centering
\includegraphics[width=0.95\textwidth]{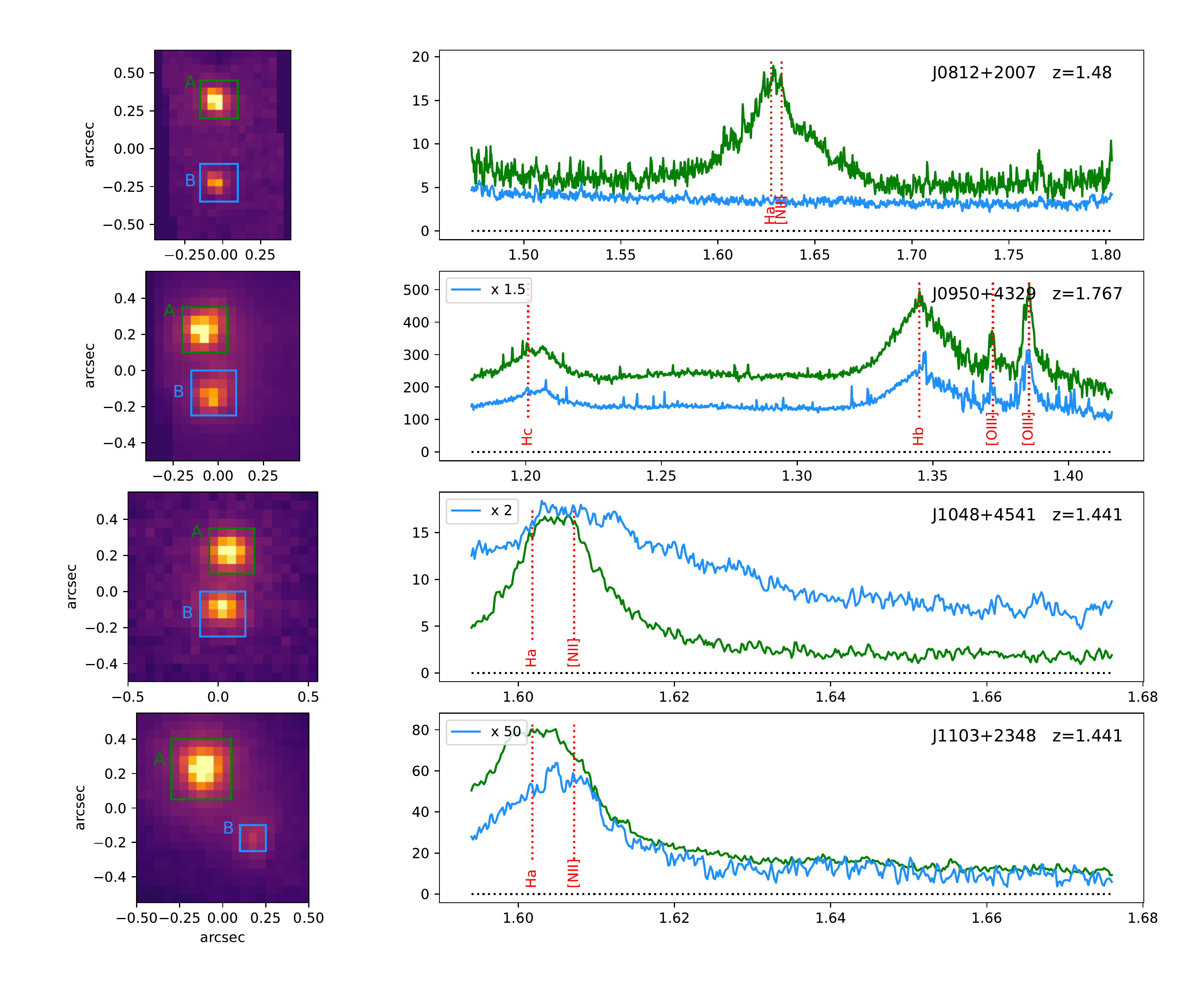}
\caption{H$\alpha$ emission line maps (left) and spectra (right) of the systems observed with OSIRIS (target name and redshift reported in the right panels). The line maps are oriented with North to the top and West to the right. The spectra shown on the right panels have been extracted over the squared apertures marked on the left panes (with the same color-coding). Each component of the systems is labelled as in Table~\ref{tab:results}. To optimize the visualization, some of the spectra have been multiplied by arbitrary factors, which are reported in the top-left corner of the corresponding panels. Vertical dotted lines show the position of the main expected emission lines.
    }
\label{fig:spectra}
\end{figure*}

\begin{table*}[]
\begin{center}
\begin{tabular}{ccccc}
\hline
\hline
Target    & Instrument & Band & T$_{exp}$(min) & FWHM($\arcsec$)  \\
\hline
J0812+2007	& OSIRIS & Hbb  &  60  & 0.11 \\
J0950+4329	& OSIRIS & Jbb  &  30  & 0.15 \\
J1048+4541	& OSIRIS & Hn3  &  45  & 0.14 \\
J1103+2348	& OSIRIS & Hn3  &  75 & 0.17 \\
J1318--0136  & ERIS   & Hlow & 40 & 0.12 \\
\hline
\end{tabular}
\end{center}
\caption{
Main properties of the five targets observed with OSIRIS/Keck and ERIS/VLT, together with the observational setup. Redshift are as reported in the Milliquas catalog. Bands are reported as in the instrument user 
manuals
The table also reports the FWHM of the PSF, calculated on isolated sources.
}
\label{tab:AOspectra}
\end{table*}

\subsection{Classification using unresolved spectroscopy }
\label{sec:deconv}

Ground-based spectra are primarily used to discover AGNs and measure their redshifts.  Most of the available QSO spectra come from the SDSS DR16Q \citep{Lyke20}, 2QZ \citep{Croom04} and LAMOST \citep{Yao19} surveys. As shown in Sec.~\ref{sec:effic}, in GMP-selected systems the luminosity ratio between the primary and the secondary object is usually not large, and it is below a factor of six for all objects with G$>$18.5. For this reason, ground-based spectra, which are usually blend of all the components, can also be used to identify the systems containing a star if they can be reproduced with the superposition of an AGN and a star.

Following the procedure described in \cite{Scialpi23}, we have performed the deconvolution of the spectra of all the targets in Tab.~\ref{tab:summary}. The observed spectra are fitted with a combination of an AGN \citep{Vandenberk01,Temple2021} and a stellar template \citep{Covey07} . The fitting parameters, determined with a $\chi^2$ minimization, are the normalization of the two spectra, the redshift of the AGN, the radial velocity of the star (limited to $\pm300$km/sec), and the level of dust extinction of the AGN template.  LBT observations were obtained without performing any spectroscopic analysis, thus the LBT sample is unbiased towards systems constituted by 2 AGNs (either dual or lensed) and AGN/star associations. In contrast, Keck observations were targeted towards systems with no clear stellar features in the blended ground-based spectra, except J0812+2007 that was observed because already included in the LBT sample. Also the ERIS target J1318-0136 was observed because part of the LBT sample.

The results of the deconvolution for the objects in Table~\ref{tab:summary} are reported in Table~\ref{tab:comparison} and plotted in the appendix. 
They show that:
\begin{itemize}
\item  J0732+3533, J0950+4329, J1048+4541, and J1103+2348 can be fitted well by a single AGN spectrum and do not show any stellar feature. 

\item In contrast, J0812+2007, J0812-0040 and J1510+5959 show clear stellar features, revealing the presence of a projected star. 

\item J1318-0136 has a low-S/N, uncalibrated spectrum from 2QZ survey \citep{Croom04}, with problematic features. Our estimate of the redshift is based on the presence of a line at 
$\sim3850\AA$, identified as CIV$]\lambda$1549 based on the presence of two low-significance features at 4800 \AA and 6950 \AA, possibly identified as CIII and MgII. Even if our estimate is in agreement with that originally provided by 2QZ, the redshift determination is not certain. Also, the quality of the spectrum does not allow us to look for stellar features, therefore this total spectrum does not provide any classification. 

\item J0927+3512  is a peculiar object. It was observed by LAMOST and classified as a QSO at z=1.149 based on a broad feature at 4050 \AA, nevertheless this feature has low S/N and no other convincing emission line is present. Also, the MgII line expected at $\sim6000$\AA for this redshift is not present, and the continuum shape shows a peculiar shape, with a minimum in $F_\lambda$ around 7500\AA. This spectrum does not convincingly show that an AGN is at all present in this system, therefore J0927+3512 remains unclassified.
\end{itemize}

The appendix reports the spectra and their spectral deconvolution. The best fit is shown in red, the QSO in blue, the stellar emission in orange, with pink vertical stripes of width $50\AA$ covering the main stellar absorption features of the best-fitting stellar type. 

\subsection{Classification using resolved spectroscopy}
\label{sec:AOspectra}

We observed four GMP-selected systems with the integral-field spectrograph (IFS) OSIRIS on the Keck telescope, and one target with ERIS/SPIFFIER on the VLT. On both cases the use of laser-assisted adaptive optics allowed us to fully resolve the systems and obtain independent spectra for the two components. 
These sources were selected to have \fmp$\geq10$, a more conservative value then the minimum \fmp=8 derived in Sec.~\ref{sec:effic}.
We extracted the targets from the Milliquas v7.9 catalog \citep{Flesch21},  considering known AGNs with a spectroscopic redshift such as to put either H$\alpha$ or H$\beta$ inside one of the available 
near-IR bands 
($0.85<z<1.11$, $1.28<z<1.85$, and $2.03<z<2.65$ for OSIRIS; 
$0.70<z<1.02$, $1.27<z<1.73$, and $1.98<z<2.73$ for ERIS).
We selected targets at high galactic latitude ($b>20$\degree) and near a star bright enough to drive the AO system. 
We selected five targets, three of which
(J0812+2007, J0950+4329, and J1318-0136) 
are in the LBT sample described in Sec.~\ref{sec:lbt}.
Table~\ref{tab:AOspectra} summarizes the obtained observations, which are described in the next sections. 

\subsubsection{Keck/OSIRIS IFU  spectra}
\label{sec:keck}

Keck spectra were obtained on on February 8th, 2023, using a pixel scale of 50~mas.
In all cases the AO module was driven by a laser guide star (LGS) and a nearby natural tip-tilt (TT) star to correct the low orders.
The positions of the two components in J0812+2007 and J0950+4329 were known before the observations from the Gaia catalog and the LBT observations. The knowledge of the position angle in these two cases allowed us to use a smaller field-of-view (0.8$\arcsec\times3.2\arcsec$) with a broader wavelength coverage (OSIRIS filters Hbb and Jbb, respectively). 
In contrast, J1048+4541 and J1103+2348 are isolated sources in the Gaia database and the position of the companion was not known before the observations. As a consequence, we used a narrower
wavelength coverage (Hn3 on both cases) to have a larger field-of-view (1.6$\arcsec\times3.2\arcsec$), well matched to the expected separations below $0.7\arcsec$. In addition to the science targets, each night we also observed a standard star of spectral type A for telluric calibration.
All data cubes were assembled and reduced using the standard OSIRIS pipeline \citep{Lockhart19}, see \cite{Ciurlo23} for more details.
Fig.~\ref{fig:spectra} shows the 
images of the observed systems (left panels) and the spectra (right panels) obtained via a weighted sum in the shown apertures.

\begin{table*}[]
\centering
\begin{tabular}{clcclccc}
\hline
\hline
Target      &  Class         & \mc{Separation}  & Lines        & Center & FWHM & redshift  \\
            &                & arcsec  & kpc    &              & ($\mu$m) & (km/s) & \\
\hline
J0812+2007A & AGN            &\mr{0.61}&\mr{-}  & H$\alpha$    & 1.628  &  7400 &  1.481 \\     
J0812+2007B & Star           &         &       &   -          &   -     &   -  &  - \\
\hline
J0950+4329A & \mr{lensed AGN}&\mr{0.37}&\mr{2.0}&H$\alpha$+[NII]& 1.345 & 4890 & 1.767\\
J0950+4329B &                &         &        &H$\alpha$+[NII]& 1.345 & 4830 & 1.767\\
\hline
J1048+4541A & \mr{dual AGN}  &\mr{0.30}&\mr{2.6}&H$\beta$       & 1.605 & 2520 & 1.446\\
J1048+4541B &                &         &        &H$\beta$       & 1.608 & 5330 & 1.450\\
\hline
J1103+2348A &\mr{dual AGN}   &\mr{0.52}&\mr{4.4}&H$\alpha$+[NII]& 1.602 & 2990 & 1.442 \\
J1103+2348B &                &         &        &H$\alpha$+[NII]& 1.605 & 2960 & 1.445 \\
\hline

J1318-0136A & AGN            &\mr{0.27}&\mr{-}  & H$\alpha$     & 1.637 & 2990 & 1.495 \\           
J1318-0136B & Star           &         &        &        -      & -      &   -  &   -   \\
\hline
\end{tabular}
\caption{
    Summary of the results from VLT?ERIS and Keck/OSIRIS observations: most probable classification, projected angular and linear distances from the brightest object, and center of the observed lines.
}
\label{tab:results}
\end{table*}

All the systems show the presence of two point-like, unresolved sources (see Fig.~\ref{fig:spectra}). 
J0812+2007 is constituted by an AGN, revealed by the broad \Ha\ line, and a second object characterized by a featureless continuum. The most probable interpretation for the latter is a star, which is expected in about 30\% of the systems \citep{Mannucci22}.
Both components of J0950+4320 show QSO spectra, 
characterized by broad \Hb\ and \Hc\ and bright [OIII]4959,5007 lines. 
The two spectra become virtually identical by applying a small correction for dust extinction to component A for $A_V=0.4$, and a normalization factor of 2.65.   The similarity of the spectra and the luminosity of the object (see Fig.~\ref{fig:NIRcolors}) suggest that this is a lensed QSO, with the different dust extinction produced by the different light paths for the two images. The lensing galaxy is not detected but this is compatible with the relatively low sensitivity of our spectra to extended and continuum-dominated objects close to two bright point sources.
Both J1048+4541 and J1103+2348 are constituted by two components each, and all the spectra show the presence of bright, very broad \Ha\ lines. The line profiles in each pair are very different, both in center and shapes. To quantify the difference, we fit a Gaussian line and a constant continuum to each broad emission line. The formal errors on both line center and FWHM are small, a few km/sec, but the real uncertainties, due to the non-Gaussian profiles of the line and the partial coverage of some of the lines, are larger. For this reason we estimate typical uncertainties of $\sim$100-200~km/sec on the centers and 200-500~km/sec on the FWHM. The  \Ha\ central wavelengths of the two components of J1048+4541 differ by $500\pm300$~km/s, and the two lines have  very different line widths, i.e. FWHM = 2500$\pm200$~km/s for component A and FWHM=5300$\pm500$~km/s for component B. The significance of the velocity offset is hampered by the lack of 
most of the blue part of the line of component B, while the difference in line width is highly significant. The two components of J1103+2348 have similar line widths (FWHM=$2950\pm400$~km/s in both cases) but the central wavelengths are offset by $430\pm200$~km/s. In addition, the line shapes are very different, with opposite asymmetries. These differences are not compatible with two lensed images of one variable source but with different delays: indeed, in both cases the expected time delay between the two components is a few days at most, while the luminosities of the objects imply BLR sizes of hundreds of days \citep{Bentz13}.

\begin{figure*}
\centering
\includegraphics[width=0.25\textwidth]{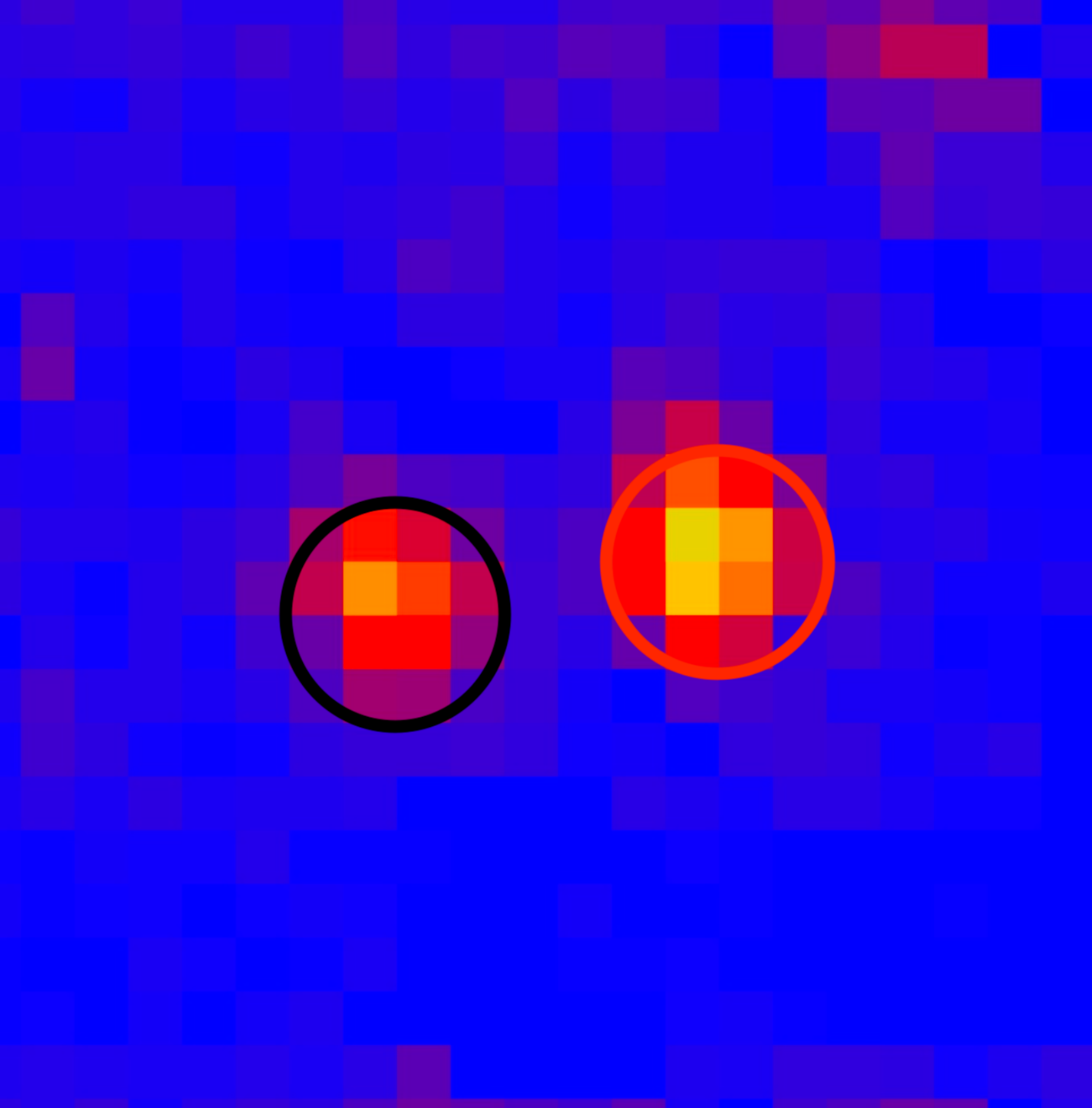}
\includegraphics[width=0.47\textwidth]{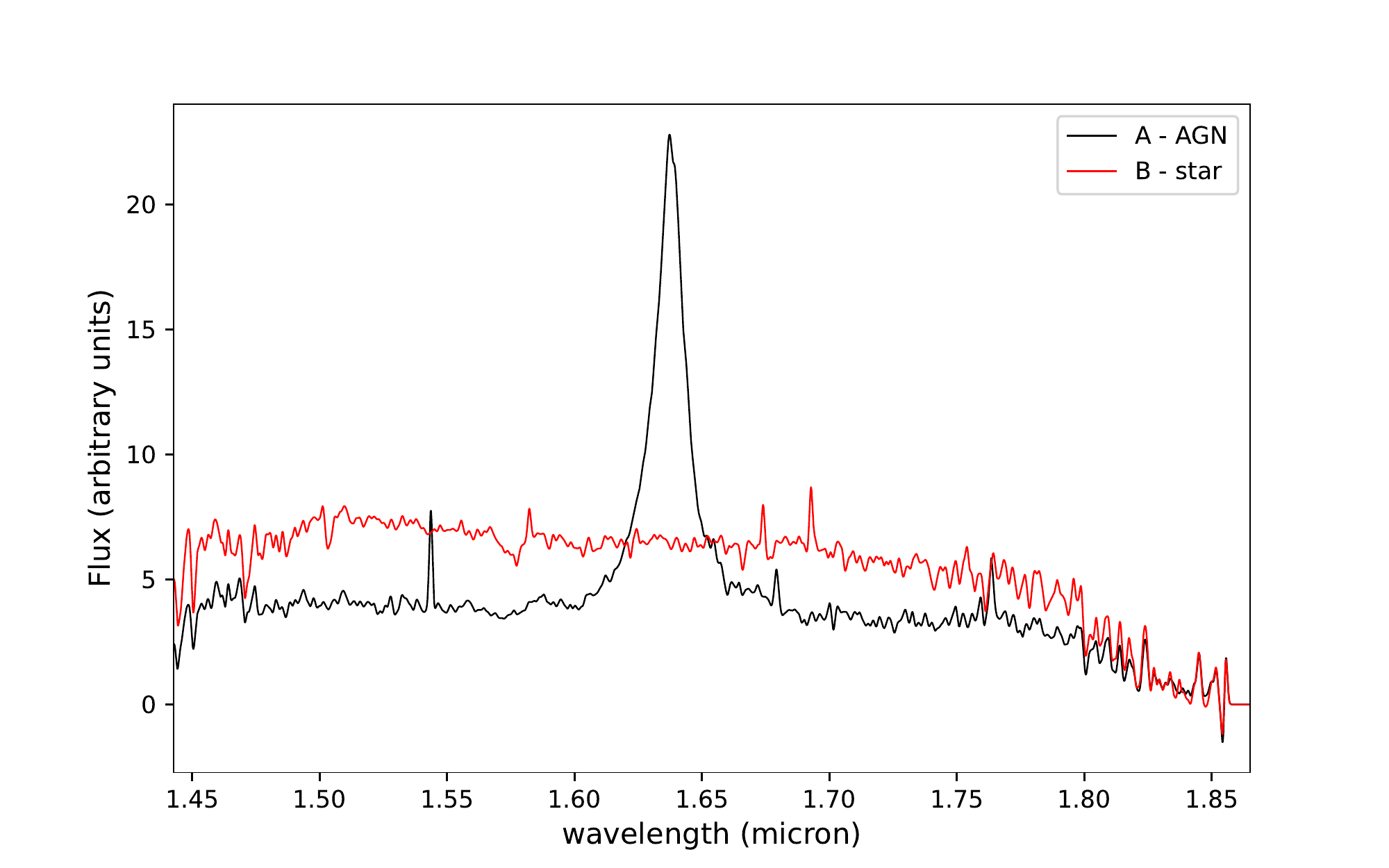}
\caption{Image (left) and uncalibrated H-band spectra (right) of  J1318-0136, obtained with VLT/ERIS. The image has $1\arcsec \times 1\arcsec$ size and is oriented with North to the top and West to the right. The spectra on the right are extracted using circular apertures of 200~mas diameter shown on the image. The bright emission line is the H$\alpha$ of AGN, while the other component only shows absorption features.
}
\label{fig:J1318-0136}
\end{figure*}

\subsubsection{VLT/ERIS IFU spectra}
\label{sec:eris}

J1318-0136 was observed on April 14th 2023 with the new Enhanced Resolution Imager and Spectrograph (ERIS, \citealt{Davies23}) on the ESO/VLT, during the first run of the INAF GTO time on this instrument, using a LGS and a nearby TT star. We used the SPIFFIER IFU spectrograph \citep{George16}, upgrade of the SPIFFI spectrograph \cite{Eisenhauer03}, covering a field of view of 3.2$\arcsec\times3.2\arcsec$ with 50~mas sampling. We integrated for 30min using the H-band grating, which provides a spectral resolution of R$\sim$5200 in the wavelength range between 1.45~$\mu$m and 1.85~$\mu$m. The data were reduced with the automatic pipeline\footnote{https://www.eso.org/sci/software/pipelines/eris/eris-pipe-recipes.html}. The PSF has FWHM=120~mas, limited by the 50~mas spaxel scale. 

In Fig. \ref{fig:J1318-0136}, we show an image of the system, showing the presence of two, well-separated components at $0.27\arcsec$ of separation, and the spectra of the two sources extracted using circular apertures of 200~mas-diameter, after removing a second-pass background in each spectral plane of the cube. 
The bright H$\alpha$ emission line centered at 1.6375 $\mu$m corresponds to redshift $z=1.495$ and is in good agreement with the tabulated value of $z=1.486$ obtained from rest-frame UV lines, clearly revealing the AGN nature of the brightest component A. On the contrary, component B shows no emission lines and a few absorption features, revealing its nature as a foreground star.\\

In conclusion, Keck and VLT spectroscopy has allowed us to reliably classify five GMP-selected QSOs: two are AGN/star projection, one is a lensed systems, and two are dual AGNs, with the properties reported in Table~\ref{tab:results}. In the next section these classifications will be compared with those obtained from the near-IR colors (Sect.~\ref{sec:colors}) and the total spectra (Sect.~\ref{sec:deconv}).

\subsection{Comparison between different classifications}
\label{sec:comparison}

\begin{table}[]
\caption[]{
Classification of all systems presented in this paper, based on different methods: 
deconvolution of integrated archival spectra,
near-IR colors (from LBT images), 
and IFU spectroscopy (with Keck or ERIS).
}
\label{tab:comparison}
\centering
\begin{tabular}{cccc}
\hline
\hline
 Target      & \multicolumn{3}{c}{Classification method}\\
             & Integr.  & Colors   & Resolved \\
             & spectrum &          & spectrum\\ 
\hline
J0732$+$3533 & AGN       & Unclass. &     - \\
J0812$-$0040 & AGN+star  & AGN+star &     - \\
J0812$+$2007 & AGN+star  & Unclass. & AGN+star\\
J0927$+$3512 & Unclass.  & Stars?   &     -    \\
J0950$+$4329 & AGN       & Lensed AGN   & Lensed AGN  \\
J1048+4541	 & AGN       &  -       & Dual AGN        \\
J1103+2348	 & AGN       &  -       & Dual AGN    \\
J1318$-$0136 & Unclass. & AGN+star  & AGN+star    \\
J1510$+$5959 & AGN+star & AGN+star  &    -    \\  
\hline
\end{tabular}
\end{table}

The three independent classification based on near-IR colors, integrated spectra, and spatially resolved spectroscopy are compared in Table~\ref{tab:comparison}. It appears that there is an excellent agreement between all available classifications: 
\begin{itemize}
\item J0812--0040 and J1510+5959 are classified as AGN/star associations by the available methods, i.e, colors and integrated spectra; 
\item J0812+2007 is an AGN/star systems based on the Keck spectrum, in agreement with the integrated spectrum. The large error on its $H-K_s$ color does not allow us to obtain a definite classification from the colors, which nevertheless are compatible with the presence of a star. 
\item The classification as lensed system of J0950+4329 is obtained from the color-magnitude diagram and is confirmed by the IFU spectrum. 
\item J1048+4541 and J1103+2348 are classified as dual AGN systems by both the integrated and the IFU spectra. 
\item The spectrum of J0732+3533 does not show any stellar contribution, while the colors are compatible with both AGN/AGN and AGN/star systems.  
\item J1318--0136
is classified as an AGN/star alignment based on its near-IR colors, while the spectrum is of low quality and cannot be used for a reliable classification. The ERIS spectra confirm this classification.
\item Finally, J0927+3512 is a peculiar system: while the LAMOST survey classifies this object as a secure QSO at $z=1.149$, our spectrum does not show any clear emission line, and the  MgII$\lambda$2800 line expected at 6017 \AA~is not detected. Thus, the presence of an AGN in this system is not secure, let alone its redshift. Both components show very blue $H-K_s$ colors that, combined with the high apparent magnitude (H$<$15.5) point toward a classification as a star pair. In contrast, its spectrum is not a clear superposition of two stars. The nature of this system remains uncertain.
\end{itemize}

Summarizing, while spatially-resolved spectra are the ultimate tool to secure a classification, combining integrated spectra, magnitudes, and colors is a useful method to classify the GMP systems with relatively high confidence.

\begin{figure*}[t]
\centering
\includegraphics[width=0.98\textwidth]{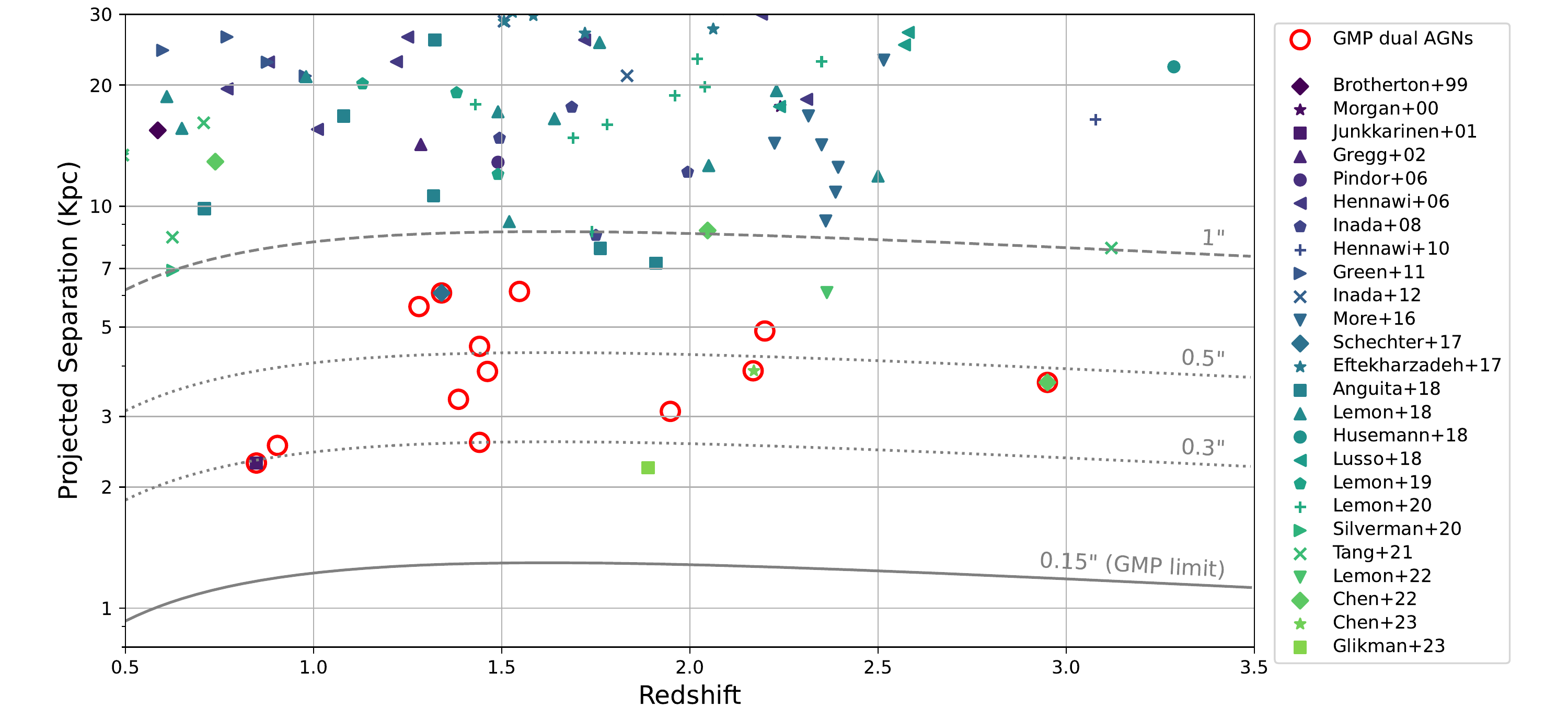}
\caption{
Projected separation vs. redshift of all the spectroscopically confirmed dual AGNs with separation below 30~kpc. Literature data are from 
\cite{Brotherton99}, \cite{Morgan00}, \cite{Junkkarinen01}, 
\cite{Gregg02}, \cite{Pindor06}, \cite{Hennawi06}, \cite{Inada08}
\cite{Hennawi10}, \cite{Green11}, \cite{Inada12}, \cite{More16},
\cite{Eftekharzadeh17}, \cite{Scheckter17}, \cite{Lemon18},
\cite{Husemann18}, 
\cite{Lusso18},
\cite{Lemon19},
\cite{Lemon20},  \cite{Silverman20}, \cite{Tang21}, \cite{Lemon22},
\cite{Chen22a}, \cite{Chen23}, and \cite{Glikman23}. 
Red circles are the GMP systems, empty circles are the dual AGNs originally selected with this techniques and described in \cite{Mannucci22}, \cite{Ciurlo23}, \cite{Scialpi23}, and this work. This data compilation can be found in computer-readable format in \url{https://tinyurl.com/dualagns}
}
\label{fig:distrib}%
\end{figure*}

\section{Discussion and conclusions}
\label{sec:summary}

In Sec.~\ref{sec:effic} we have shown that the GMP method has a simple and well defined selection function in terms of the separation of the sources and their luminosity ratios. This is a key feature to compare the properties of the discovered dual and lensed systems with the theoretical predictions, and to test, for the first time, some of the aspects of the models of hierarchical formation of galaxies and SMBH.

Based on integrated spectra and AO-assisted observations at VLT, Keck and LBT we have shown that combining integrated spectra and spatially-resolved near-IR luminosity and colors can provide indications on the nature of these systems as dual AGNs, lensed QSOs, or AGN/star alignment. 

\cite{Mannucci22}, \cite{Ciurlo23}, \cite{Scialpi23}, and this work present samples of dual AGNs selected by the GMP method and classified by resolved spectroscopy. In total, 22 GMP systems were classified, obtaining 10 dual AGNs, 5 lensed systems (including the system classified as a lens by \citealt{Li23}), and 7 AGN/star associations.  One of these systems and three more GMP candidates have been previously identified using other techniques: 
\cite{Junkkarinen01} reported the serendipitous discovery with HST of a very compact dual system at $z\sim0.84$, with two components separated by $\sim0.3\arcsec$.
\cite{Scheckter17} identified a dual system at $z=1.34$ and $\delta=0.71\arcsec$ by spatial deconvolution of ground-based images of AGNs selected to have red WISE W1--W2 colors.
\cite{Chen22a} identified several multiple AGN candidates by looking for SDSS systems associated to more than one Gaia object within a few arcsec. One of these systems was spectroscopically
confirmed to be a dual AGN at $z=2.95$ and separation $0.46\arcsec$ \citep{Mannucci22}.  As shown in Fig.~\ref{fig:effic}, this technique becomes very inefficient at $\delta<0.5\arcsec$, but sporadically can provide targets down to $\delta\sim0.3\arcsec$.  
Finally, \cite{Chen23} present a detail analysis of one target at $z=2.17$ and separation $\delta=0.46\arcsec$ selected using "varstrometry", i.e,, the Gaia astrometric extra jitter due to the unrelated variation of the two AGNs. 

Fig. \ref{fig:distrib} shows the distribution in redshift and projected separations of all the known dual AGNs with separations below 30~kpc. It is evident as all the 14 systems with separation below 7~kpc are GMP-selected systems, and that 9 of these are only detected by this method. 
The only exceptions are the complex system discovered by \cite{Lemon22} at z=2.36, which is a lensed system with 6 components with large separations (up to $\sim4\arcsec$), possible attributed to a dual systems with separation of $\sim0.7\arcsec$, a very interesting yet uncertain and peculiar systems, and
the dual system at z=1.889 and $\delta=2.2$kpc serendipitously discovered by \cite{Glikman23}, whose luminosity ratio between the components is too large to be selected by GMP (see Fig.~\ref{fig:effic2D}).

In conclusion, the GMP method appears to be the best suited technique to provide a large sample of high-redshift dual AGNs residing in the same host galaxies. We are currently in the process of confirming more systems with AO spectroscopic observations, in particular with VLT/MUSE, VLT/ERIS, and Keck/OSIRIS, to obtain a sample large enough to test the prediction of models of galaxy formation and evolution.

Summarizing: 
\begin{itemize}
\item We have tested the performances of the GMP selection using dense stellar fields with numerous projected pairs at low separations. This allowed us to estimate the level of contamination (false positives) as a function of the selecting parameter \fmp~ (Fig.~\ref{fig:effic}, left) and the selection efficiency as a function of separation $\delta$ (Fig.~\ref{fig:effic}, left) and source magnitudes
(Fig.~\ref{fig:effic2D}). The results show that low levels of contamination and high selection efficiency can be obtained for $\delta>0.15\arcsec$ and G$<$20.5;

\item AO-assisted imaging in the $H$ and $K_s$ band with SOUL/LUCI at LBT of 7 GMP-selected systems has allowed us to resolve all of them into multiple components, measure the $H-K_s$ of all the sources, and 
to develop a new color-based method  
to identify the most probable nature 
of the GMP-selected systems
(Fig.~\ref{fig:NIRcolors});

\item we reproduced the integrated, ground-based spectra of the 9 systems of this sample with a combination of an AGN template with a foreground star. 
Four of these 9 systems are well reproduced by a simple AGN spectrum
(Fig.~\ref{fig:deconv1}),
three  show clear evidence for the presence of an AGN and a star,
\ref{fig:deconv2})
and two remain unclassified , \ref{fig:deconv3}.

\item we present the spatially-resolved spectra of 5 systems with Keck/OSIRIS  (Fig.~\ref{fig:spectra}) and VLT/ERIS (Fig.~\ref{fig:J1318-0136}). These spectra has allowed us to discover two dual systems with separations of $0.30\arcsec$ and $0.52\arcsec$, a very compact lensed system (separation $0.25\arcsec$), and two AGN/star associations;

\item comparing the various classification methods, we have shown that the combined analysis of resolved colors and integrated spectra, can provide reliable classifications of the systems in most cases.

\item Fig.~\ref{fig:distrib} shows all the confirmed dual AGNs at $z>0.5$ and separations below 30~kpc. Fourteen of the fifteen systems with separations below $\sim7$~kpc are GMP systems, and 10 of them are originally selected by this method. In other words, since its introduction \citep{Mannucci22}, this techniques has already allow us to triple the confirmed systems at small separations. This result shows that the GMP technique is an important step forward to sample the dual AGNs hosted by the same galaxy and test the predictions of the models at $\delta<7$~kpc.
\end{itemize}
%

\begin{acknowledgements}

We are grateful to the teams of VLT/ERIS and LBT/SOUL for proving us with excellent instruments for this project, and to the staffs of Keck, VLT and LBT for superb support during the observations; we acknowledge the use of 2hrs of INAF discretionary time at LBT. 
We thanks Stefano Cristiani, Andrea Grazian, and 
Robert Smith for assistance with ground-based, integrated spectra.
We acknowledge financial contribution from INAF Large Grant  “Dual and binary supermassive black holes in the multi-messenger era: from galaxy mergers to gravitational waves” (Bando Ricerca Fondamentale INAF 2022), and from the French National Research Agency (grant ANR-21-CE31-0026, project MBH\_waves).\\

Based on observations collected at the European Southern Observatory under ESO programme 111.24QJ.001\\

This work has made use of data from the European Space Agency (ESA) mission
{\it Gaia} (\url{https://www.cosmos.esa.int/gaia}), processed by the {\it Gaia}
Data Processing and Analysis Consortium (DPAC,
\url{https://www.cosmos.esa.int/web/gaia/dpac/consortium}). Funding for the DPAC
has been provided by national institutions, in particular the institutions
participating in the {\it Gaia} Multilateral Agreement.\\

Funding for the Sloan Digital Sky Survey (SDSS) has been provided by the Alfred P. Sloan Foundation, the Participating Institutions, the National Aeronautics and Space Administration, the National Science Foundation, the U.S. Department of Energy, the Japanese Monbukagakusho, and the Max Planck Society. The SDSS Web site is http://www.sdss.org/. The SDSS is managed by the Astrophysical Research Consortium (ARC) for the Participating Institutions. The Participating Institutions are The University of Chicago, Fermilab, the Institute for Advanced Study, the Japan Participation Group, The Johns Hopkins University, Los Alamos National Laboratory, the Max-Planck-Institute for Astronomy (MPIA), the Max-Planck-Institute for Astrophysics (MPA), New Mexico State University, University of Pittsburgh, Princeton University, the United States Naval Observatory, and the University of Washington.

\end{acknowledgements}



\bibliographystyle{aa} 
\bibliography{references}{}

\begin{thebibliography}{68}
\expandafter\ifx\csname natexlab\endcsname\relax\def\natexlab#1{#1}\fi

\bibitem[{{Amaro-Seoane} {et~al.}(2023){Amaro-Seoane}, {Andrews}, {Arca Sedda},
  {Askar}, {Baghi}, {Balasov}, {Bartos}, {Bavera}, {Bellovary}, {Berry},
  {Berti}, {Bianchi}, {Blecha}, {Blondin}, {Bogdanovi{\'c}}, {Boissier},
  {Bonetti}, {Bonoli}, {Bortolas}, {Breivik}, {Capelo}, {Caramete},
  {Cattorini}, {Charisi}, {Chaty}, {Chen}, {Chru{\'s}li{\'n}ska}, {Chua},
  {Church}, {Colpi}, {D'Orazio}, {Danielski}, {Davies}, {Dayal}, {De Rosa},
  {Derdzinski}, {Destounis}, {Dotti}, {Du{\r{A}}{\textsterling}an}, {Dvorkin},
  {Fabj}, {Foglizzo}, {Ford}, {Fouvry}, {Franchini}, {Fragos}, {Fryer},
  {Gaspari}, {Gerosa}, {Graziani}, {Groot}, {Habouzit}, {Haggard}, {Haiman},
  {Han}, {Istrate}, {Johansson}, {Khan}, {Kimpson}, {Kokkotas}, {Kong},
  {Korol}, {Kremer}, {Kupfer}, {Lamberts}, {Larson}, {Lau}, {Liu},
  {Lloyd-Ronning}, {Lodato}, {Lupi}, {Ma}, {Maccarone}, {Mandel}, {Mangiagli},
  {Mapelli}, {Mathis}, {Mayer}, {McGee}, {McKernan}, {Miller}, {Mota},
  {Mumpower}, {Nasim}, {Nelemans}, {Noble}, {Pacucci}, {Panessa},
  {Paschalidis}, {Pfister}, {Porquet}, {Quenby}, {Ricarte}, {R{\"o}pke},
  {Regan}, {Rosswog}, {Ruiter}, {Ruiz}, {Runnoe}, {Schneider}, {Schnittman},
  {Secunda}, {Sesana}, {Seto}, {Shao}, {Shapiro}, {Sopuerta}, {Stone},
  {Suvorov}, {Tamanini}, {Tamfal}, {Tauris}, {Temmink}, {Tomsick}, {Toonen},
  {Torres-Orjuela}, {Toscani}, {Tsokaros}, {Unal}, {V{\'a}zquez-Aceves},
  {Valiante}, {van Putten}, {van Roestel}, {Vignali}, {Volonteri}, {Wu},
  {Younsi}, {Yu}, {Zane}, {Zwick}, {Antonini}, {Baibhav}, {Barausse}, {Bonilla
  Rivera}, {Branchesi}, {Branduardi-Raymont}, {Burdge}, {Chakraborty},
  {Cuadra}, {Dage}, {Davis}, {de Mink}, {Decarli}, {Doneva}, {Escoffier},
  {Gandhi}, {Haardt}, {Lousto}, {Nissanke}, {Nordhaus}, {O'Shaughnessy},
  {Portegies Zwart}, {Pound}, {Schussler}, {Sergijenko}, {Spallicci},
  {Vernieri}, \& {Vigna-G{\'o}mez}}]{Amaro-Seoane23}
{Amaro-Seoane}, P., {Andrews}, J., {Arca Sedda}, M., {et~al.} 2023, Living
  Reviews in Relativity, 26, 2

\bibitem[{{Arzoumanian} {et~al.}(2018){Arzoumanian}, {Baker}, {Brazier},
  {Burke-Spolaor}, {Chamberlin}, {Chatterjee}, {Christy}, {Cordes}, {Cornish},
  {Crawford}, {Thankful Cromartie}, {Crowter}, {DeCesar}, {Demorest}, {Dolch},
  {Ellis}, {Ferdman}, {Ferrara}, {Folkner}, {Fonseca}, {Garver-Daniels},
  {Gentile}, {Haas}, {Hazboun}, {Huerta}, {Islo}, {Jones}, {Jones}, {Kaplan},
  {Kaspi}, {Lam}, {Lazio}, {Levin}, {Lommen}, {Lorimer}, {Luo}, {Lynch},
  {Madison}, {McLaughlin}, {McWilliams}, {Mingarelli}, {Ng}, {Nice}, {Park},
  {Pennucci}, {Pol}, {Ransom}, {Ray}, {Rasskazov}, {Siemens}, {Simon},
  {Spiewak}, {Stairs}, {Stinebring}, {Stovall}, {Swiggum}, {Taylor},
  {Vallisneri}, {van Haasteren}, {Vigeland}, {Zhu}, \& {NANOGrav
  Collaboration}}]{Arzoumanian18}
{Arzoumanian}, Z., {Baker}, P.~T., {Brazier}, A., {et~al.} 2018, \apj, 859, 47

\bibitem[{{Bentz} {et~al.}(2013){Bentz}, {Denney}, {Grier}, {Barth},
  {Peterson}, {Vestergaard}, {Bennert}, {Canalizo}, {De Rosa}, {Filippenko},
  {Gates}, {Greene}, {Li}, {Malkan}, {Pogge}, {Stern}, {Treu}, \&
  {Woo}}]{Bentz13}
{Bentz}, M.~C., {Denney}, K.~D., {Grier}, C.~J., {et~al.} 2013, \apj, 767, 149

\bibitem[{{Brotherton} {et~al.}(1999){Brotherton}, {Gregg}, {Becker},
  {Laurent-Muehleisen}, {White}, \& {Stanford}}]{Brotherton99}
{Brotherton}, M.~S., {Gregg}, M.~D., {Becker}, R.~H., {et~al.} 1999, The
  Astrophysical Journal, 514, L61

\bibitem[{{Capelo} {et~al.}(2017){Capelo}, {Dotti}, {Volonteri}, {Mayer},
  {Bellovary}, \& {Shen}}]{Capelo2017}
{Capelo}, P.~R., {Dotti}, M., {Volonteri}, M., {et~al.} 2017, \mnras, 469, 4437

\bibitem[{{Chen} {et~al.}(2022{\natexlab{a}}){Chen}, {Di Matteo}, {Ni},
  {Tremmel}, {DeGraf}, {Shen}, {Holgado}, {Bird}, {Croft}, \& {Feng}}]{Chen22c}
{Chen}, N., {Di Matteo}, T., {Ni}, Y., {et~al.} 2022{\natexlab{a}}, arXiv
  e-prints, arXiv:2208.04970

\bibitem[{{Chen}(2021)}]{Chen22b}
{Chen}, Y.-C. 2021, arXiv e-prints, arXiv:2109.06881

\bibitem[{{Chen} {et~al.}(2022{\natexlab{b}}){Chen}, {Hwang}, {Shen}, {Liu},
  {Zakamska}, {Yang}, \& {Li}}]{Chen22a}
{Chen}, Y.-C., {Hwang}, H.-C., {Shen}, Y., {et~al.} 2022{\natexlab{b}}, \apj,
  925, 162

\bibitem[{{Chen} {et~al.}(2023){Chen}, {Liu}, {Foord}, {Shen}, {Oguri}, {Chen},
  {Di Matteo}, {Holgado}, {Hwang}, \& {Zakamska}}]{Chen23}
{Chen}, Y.-C., {Liu}, X., {Foord}, A., {et~al.} 2023, \nat, 616, 45

\bibitem[{{Ciurlo} {et~al.}(2023){Ciurlo}, {Mannucci}, {Yeh}, {Amiri},
  {Carniani}, {Cicone}, {Cresci}, {Khatun}, {Lusso}, {Marasco}, {Marconcini},
  {Marconi}, {Nardini}, {Pancino}, {Rosati}, {Severgnini}, {Scialpi}, {Tozzi},
  {Venturi}, {Vignali}, \& {Volonteri}}]{Ciurlo23}
{Ciurlo}, A., {Mannucci}, F., {Yeh}, S., {et~al.} 2023, arXiv e-prints,
  arXiv:2301.03091

\bibitem[{{Civano} {et~al.}(2016){Civano}, {Marchesi}, {Comastri}, {Urry},
  {Elvis}, {Cappelluti}, {Puccetti}, {Brusa}, {Zamorani}, {Hasinger},
  {Aldcroft}, {Alexander}, {Allevato}, {Brunner}, {Capak}, {Finoguenov},
  {Fiore}, {Fruscione}, {Gilli}, {Glotfelty}, {Griffiths}, {Hao}, {Harrison},
  {Jahnke}, {Kartaltepe}, {Karim}, {LaMassa}, {Lanzuisi}, {Miyaji}, {Ranalli},
  {Salvato}, {Sargent}, {Scoville}, {Schawinski}, {Schinnerer}, {Silverman},
  {Smolcic}, {Stern}, {Toft}, {Trakhtenbrot}, {Treister}, \&
  {Vignali}}]{Civano16}
{Civano}, F., {Marchesi}, S., {Comastri}, A., {et~al.} 2016, \apj, 819, 62

\bibitem[{{Covey} {et~al.}(2007){Covey}, {Ivezi{\'c}}, {Schlegel},
  {Finkbeiner}, {Padmanabhan}, {Lupton}, {Ag{\"u}eros}, {Bochanski}, {Hawley},
  {West}, {Seth}, {Kimball}, {Gogarten}, {Claire}, {Haggard}, {Kaib},
  {Schneider}, \& {Sesar}}]{Covey07}
{Covey}, K.~R., {Ivezi{\'c}}, {\v{Z}}., {Schlegel}, D., {et~al.} 2007, \aj,
  134, 2398

\bibitem[{{Croom} {et~al.}(2009){Croom}, {Richards}, {Shanks}, {Boyle},
  {Strauss}, {Myers}, {Nichol}, {Pimbblet}, {Ross}, {Schneider}, {Sharp}, \&
  {Wake}}]{Croom09}
{Croom}, S.~M., {Richards}, G.~T., {Shanks}, T., {et~al.} 2009, \mnras, 399,
  1755

\bibitem[{{Croom} {et~al.}(2004){Croom}, {Smith}, {Boyle}, {Shanks}, {Miller},
  {Outram}, \& {Loaring}}]{Croom04}
{Croom}, S.~M., {Smith}, R.~J., {Boyle}, B.~J., {et~al.} 2004, \mnras, 349,
  1397

\bibitem[{{Davies} {et~al.}(2023){Davies}, {Absil}, {Agapito}, {Agudo Berbel},
  {Baruffolo}, {Biliotti}, {Bonaglia}, {Bonse}, {Briguglio}, {Campana}, {Cao},
  {Carbonaro}, {Cortes}, {Cresci}, {Dallilar}, {Dannert}, {De Rosa},
  {Deysenroth}, {Di Antonio}, {Di Cianno}, {Di Rico}, {Doelman}, {Dolci},
  {Dorn}, {Eisenhauer}, {Esposito}, {Fantinel}, {Ferruzzi}, {Feuchtgruber},
  {F{\"o}rster Schreiber}, {Gao}, {Gemperlein}, {Genzel}, {Gillessen},
  {Ginski}, {Glauser}, {Glindemann}, {Grani}, {Hartl}, {Hayoz}, {Heida},
  {Henry}, {Huber}, {Kasper}, {Keller}, {Kenworthy}, {Kravchenko},
  {Kuntschner}, {Lacour}, {Lightfoot}, {Lunney}, {Macintosh}, {Mannucci},
  {Marsset}, {Modigliani}, {Neeser}, {Orban de Xivry}, {Pallanca}, {Patapis},
  {Pearson}, {Pe{\~n}a}, {Percheron}, {Puglisi}, {Quanz}, {Rau}, {Riccardi},
  {Salasnich}, {Schmid}, {Serra}, {Snik}, {Sturm}, {Taylor}, {Valentini},
  {Waring}, {Wiezorrek}, \& {Xompero}}]{Davies23}
{Davies}, R., {Absil}, O., {Agapito}, G., {et~al.} 2023, arXiv e-prints,
  arXiv:2304.02343

\bibitem[{{De Rosa} {et~al.}(2019){De Rosa}, {Vignali}, {Bogdanovi{\'c}},
  {Capelo}, {Charisi}, {Dotti}, {Husemann}, {Lusso}, {Mayer}, {Paragi},
  {Runnoe}, {Sesana}, {Steinborn}, {Bianchi}, {Colpi}, {del Valle}, {Frey},
  {Gab{\'a}nyi}, {Giustini}, {Guainazzi}, {Haiman}, {Herrera Ruiz},
  {Herrero-Illana}, {Iwasawa}, {Komossa}, {Lena}, {Loiseau}, {Perez-Torres},
  {Piconcelli}, \& {Volonteri}}]{derosa20}
{De Rosa}, A., {Vignali}, C., {Bogdanovi{\'c}}, T., {et~al.} 2019, \nar, 86,
  101525

\bibitem[{{DeGraf} {et~al.}(2023){DeGraf}, {Chen}, {Ni}, {Di Matteo}, {Bird},
  {Tremmel}, \& {Croft}}]{DeGraf23}
{DeGraf}, C., {Chen}, N., {Ni}, Y., {et~al.} 2023, arXiv e-prints,
  arXiv:2302.00702

\bibitem[{{Dong-P{\'a}ez} {et~al.}(2023){Dong-P{\'a}ez}, {Volonteri},
  {Beckmann}, {Dubois}, {Trebitsch}, {Mangiagli}, {Vergani}, \&
  {Webb}}]{Dong-Paez23}
{Dong-P{\'a}ez}, C.~A., {Volonteri}, M., {Beckmann}, R.~S., {et~al.} 2023,
  arXiv e-prints, arXiv:2303.00766

\bibitem[{{Eftekharzadeh} {et~al.}(2017){Eftekharzadeh}, {Myers}, {Hennawi},
  {Djorgovski}, {Richards}, {Mahabal}, \& {Graham}}]{Eftekharzadeh17}
{Eftekharzadeh}, S., {Myers}, A.~D., {Hennawi}, J.~F., {et~al.} 2017, \mnras,
  468, 77

\bibitem[{{Eisenhauer} {et~al.}(2003){Eisenhauer}, {Abuter}, {Bickert},
  {Biancat-Marchet}, {Bonnet}, {Brynnel}, {Conzelmann}, {Delabre}, {Donaldson},
  {Farinato}, {Fedrigo}, {Genzel}, {Hubin}, {Iserlohe}, {Kasper},
  {Kissler-Patig}, {Monnet}, {Roehrle}, {Schreiber}, {Stroebele}, {Tecza},
  {Thatte}, \& {Weisz}}]{Eisenhauer03}
{Eisenhauer}, F., {Abuter}, R., {Bickert}, K., {et~al.} 2003, in Society of
  Photo-Optical Instrumentation Engineers (SPIE) Conference Series, Vol. 4841,
  Instrument Design and Performance for Optical/Infrared Ground-based
  Telescopes, ed. M.~{Iye} \& A.~F.~M. {Moorwood}, 1548--1561

\bibitem[{{Euclid Collaboration} {et~al.}(2022){Euclid Collaboration},
  {Scaramella}, {Amiaux}, {Mellier}, {Burigana}, {Carvalho}, {Cuillandre}, {Da
  Silva}, {Derosa}, {Dinis}, {Maiorano}, {Maris}, {Tereno}, {Laureijs},
  {Boenke}, {Buenadicha}, {Dupac}, {Gaspar Venancio}, {G{\'o}mez-{\'A}lvarez},
  {Hoar}, {Lorenzo Alvarez}, {Racca}, {Saavedra-Criado}, {Schwartz}, {Vavrek},
  {Schirmer}, {Aussel}, {Azzollini}, {Cardone}, {Cropper}, {Ealet}, {Garilli},
  {Gillard}, {Granett}, {Guzzo}, {Hoekstra}, {Jahnke}, {Kitching}, {Maciaszek},
  {Meneghetti}, {Miller}, {Nakajima}, {Niemi}, {Pasian}, {Percival},
  {Pottinger}, {Sauvage}, {Scodeggio}, {Wachter}, {Zacchei}, {Aghanim},
  {Amara}, {Auphan}, {Auricchio}, {Awan}, {Balestra}, {Bender}, {Bodendorf},
  {Bonino}, {Branchini}, {Brau-Nogue}, {Brescia}, {Candini}, {Capobianco},
  {Carbone}, {Carlberg}, {Carretero}, {Casas}, {Castander}, {Castellano},
  {Cavuoti}, {Cimatti}, {Cledassou}, {Congedo}, {Conselice}, {Conversi},
  {Copin}, {Corcione}, {Costille}, {Courbin}, {Degaudenzi}, {Douspis},
  {Dubath}, {Duncan}, {Dusini}, {Farrens}, {Ferriol}, {Fosalba}, {Fourmanoit},
  {Frailis}, {Franceschi}, {Franzetti}, {Fumana}, {Gillis}, {Giocoli},
  {Grazian}, {Grupp}, {Haugan}, {Holmes}, {Hormuth}, {Hudelot}, {Kermiche},
  {Kiessling}, {Kilbinger}, {Kohley}, {Kubik}, {K{\"u}mmel}, {Kunz},
  {Kurki-Suonio}, {Lahav}, {Ligori}, {Lilje}, {Lloro}, {Mansutti}, {Marggraf},
  {Markovic}, {Marulli}, {Massey}, {Maurogordato}, {Melchior}, {Merlin},
  {Meylan}, {Mohr}, {Moresco}, {Morin}, {Moscardini}, {Munari}, {Nichol},
  {Padilla}, {Paltani}, {Peacock}, {Pedersen}, {Pettorino}, {Pires}, {Poncet},
  {Popa}, {Pozzetti}, {Raison}, {Rebolo}, {Rhodes}, {Rix}, {Roncarelli},
  {Rossetti}, {Saglia}, {Schneider}, {Schrabback}, {Secroun}, {Seidel},
  {Serrano}, {Sirignano}, {Sirri}, {Skottfelt}, {Stanco}, {Starck},
  {Tallada-Cresp{\'\i}}, {Tavagnacco}, {Taylor}, {Teplitz}, {Toledo-Moreo},
  {Torradeflot}, {Trifoglio}, {Valentijn}, {Valenziano}, {Verdoes Kleijn},
  {Wang}, {Welikala}, {Weller}, {Wetzstein}, {Zamorani}, {Zoubian}, {Andreon},
  {Baldi}, {Bardelli}, {Boucaud}, {Camera}, {Di Ferdinando}, {Fabbian},
  {Farinelli}, {Galeotta}, {Graci{\'a}-Carpio}, {Maino}, {Medinaceli}, {Mei},
  {Neissner}, {Polenta}, {Renzi}, {Romelli}, {Rosset}, {Sureau}, {Tenti},
  {Vassallo}, {Zucca}, {Baccigalupi}, {Balaguera-Antol{\'\i}nez}, {Battaglia},
  {Biviano}, {Borgani}, {Bozzo}, {Cabanac}, {Cappi}, {Casas}, {Castignani},
  {Colodro-Conde}, {Coupon}, {Courtois}, {Cuby}, {de la Torre}, {Desai},
  {Dole}, {Fabricius}, {Farina}, {Ferreira}, {Finelli}, {Flose-Reimberg},
  {Fotopoulou}, {Ganga}, {Gozaliasl}, {Hook}, {Keihanen}, {Kirkpatrick},
  {Liebing}, {Lindholm}, {Mainetti}, {Martinelli}, {Martinet}, {Maturi},
  {McCracken}, {Metcalf}, {Morgante}, {Nightingale}, {Nucita}, {Patrizii},
  {Potter}, {Riccio}, {S{\'a}nchez}, {Sapone}, {Schewtschenko}, {Schultheis},
  {Scottez}, {Teyssier}, {Tutusaus}, {Valiviita}, {Viel}, {Vriend}, \&
  {Whittaker}}]{Scaramella22}
{Euclid Collaboration}, {Scaramella}, R., {Amiaux}, J., {et~al.} 2022, \aap,
  662, A112

\bibitem[{{Flesch}(2021)}]{Flesch21}
{Flesch}, E.~W. 2021, arXiv e-prints, arXiv:2105.12985

\bibitem[{{Gaia Collaboration} {et~al.}(2016){Gaia Collaboration}, {Prusti},
  {de Bruijne}, {Brown}, {Vallenari}, {Babusiaux}, {Bailer-Jones}, {Bastian},
  {Biermann}, {Evans}, {Eyer}, {Jansen}, {Jordi}, {Klioner}, {Lammers},
  {Lindegren}, {Luri}, {Mignard}, {Milligan}, {Panem}, {Poinsignon},
  {Pourbaix}, {Randich}, {Sarri}, {Sartoretti}, {Siddiqui}, {Soubiran},
  {Valette}, {van Leeuwen}, {Walton}, {Aerts}, {Arenou}, {Cropper}, {Drimmel},
  {H{\o}g}, {Katz}, {Lattanzi}, {O'Mullane}, {Grebel}, {Holland}, {Huc},
  {Passot}, {Bramante}, {Cacciari}, {Casta{\~n}eda}, {Chaoul}, {Cheek}, {De
  Angeli}, {Fabricius}, {Guerra}, {Hern{\'a}ndez}, {Jean-Antoine-Piccolo},
  {Masana}, {Messineo}, {Mowlavi}, {Nienartowicz}, {Ord{\'o}{\~n}ez-Blanco},
  {Panuzzo}, {Portell}, {Richards}, {Riello}, {Seabroke}, {Tanga},
  {Th{\'e}venin}, {Torra}, {Els}, {Gracia-Abril}, {Comoretto},
  {Garcia-Reinaldos}, {Lock}, {Mercier}, {Altmann}, {Andrae}, {Astraatmadja},
  {Bellas-Velidis}, {Benson}, {Berthier}, {Blomme}, {Busso}, {Carry},
  {Cellino}, {Clementini}, {Cowell}, {Creevey}, {Cuypers}, {Davidson}, {De
  Ridder}, {de Torres}, {Delchambre}, {Dell'Oro}, {Ducourant}, {Fr{\'e}mat},
  {Garc{\'\i}a-Torres}, {Gosset}, {Halbwachs}, {Hambly}, {Harrison}, {Hauser},
  {Hestroffer}, {Hodgkin}, {Huckle}, {Hutton}, {Jasniewicz}, {Jordan},
  {Kontizas}, {Korn}, {Lanzafame}, {Manteiga}, {Moitinho}, {Muinonen},
  {Osinde}, {Pancino}, {Pauwels}, {Petit}, {Recio-Blanco}, {Robin}, {Sarro},
  {Siopis}, {Smith}, {Smith}, {Sozzetti}, {Thuillot}, {van Reeven}, {Viala},
  {Abbas}, {Abreu Aramburu}, {Accart}, {Aguado}, {Allan}, {Allasia},
  {Altavilla}, {{\'A}lvarez}, {Alves}, {Anderson}, {Andrei}, {Anglada Varela},
  {Antiche}, {Antoja}, {Ant{\'o}n}, {Arcay}, {Atzei}, {Ayache}, {Bach},
  {Baker}, {Balaguer-N{\'u}{\~n}ez}, {Barache}, {Barata}, {Barbier}, {Barblan},
  {Baroni}, {Barrado y Navascu{\'e}s}, {Barros}, {Barstow}, {Becciani},
  {Bellazzini}, {Bellei}, {Bello Garc{\'\i}a}, {Belokurov}, {Bendjoya},
  {Berihuete}, {Bianchi}, {Bienaym{\'e}}, {Billebaud}, {Blagorodnova},
  {Blanco-Cuaresma}, {Boch}, {Bombrun}, {Borrachero}, {Bouquillon}, {Bourda},
  {Bouy}, {Bragaglia}, {Breddels}, {Brouillet}, {Br{\"u}semeister},
  {Bucciarelli}, {Budnik}, {Burgess}, {Burgon}, {Burlacu}, {Busonero}, {Buzzi},
  {Caffau}, {Cambras}, {Campbell}, {Cancelliere}, {Cantat-Gaudin}, {Carlucci},
  {Carrasco}, {Castellani}, {Charlot}, {Charnas}, {Charvet}, {Chassat},
  {Chiavassa}, {Clotet}, {Cocozza}, {Collins}, {Collins}, {Costigan}, {Crifo},
  {Cross}, {Crosta}, {Crowley}, {Dafonte}, {Damerdji}, {Dapergolas}, {David},
  {David}, {De Cat}, {de Felice}, {de Laverny}, {De Luise}, {De March}, {de
  Martino}, {de Souza}, {Debosscher}, {del Pozo}, {Delbo}, {Delgado},
  {Delgado}, {di Marco}, {Di Matteo}, {Diakite}, {Distefano}, {Dolding}, {Dos
  Anjos}, {Drazinos}, {Dur{\'a}n}, {Dzigan}, {Ecale}, {Edvardsson}, {Enke},
  {Erdmann}, {Escolar}, {Espina}, {Evans}, {Eynard Bontemps}, {Fabre},
  {Fabrizio}, {Faigler}, {Falc{\~a}o}, {Farr{\`a}s Casas}, {Faye}, {Federici},
  {Fedorets}, {Fern{\'a}ndez-Hern{\'a}ndez}, {Fernique}, {Fienga}, {Figueras},
  {Filippi}, {Findeisen}, {Fonti}, {Fouesneau}, {Fraile}, {Fraser}, {Fuchs},
  {Furnell}, {Gai}, {Galleti}, {Galluccio}, {Garabato}, {Garc{\'\i}a-Sedano},
  {Gar{\'e}}, {Garofalo}, {Garralda}, {Gavras}, {Gerssen}, {Geyer}, {Gilmore},
  {Girona}, {Giuffrida}, {Gomes}, {Gonz{\'a}lez-Marcos},
  {Gonz{\'a}lez-N{\'u}{\~n}ez}, {Gonz{\'a}lez-Vidal}, {Granvik}, {Guerrier},
  {Guillout}, {Guiraud}, {G{\'u}rpide}, {Guti{\'e}rrez-S{\'a}nchez}, {Guy},
  {Haigron}, {Hatzidimitriou}, {Haywood}, {Heiter}, {Helmi}, {Hobbs},
  {Hofmann}, {Holl}, {Holland}, {Hunt}, {Hypki}, {Icardi}, {Irwin}, {Jevardat
  de Fombelle}, {Jofr{\'e}}, {Jonker}, {Jorissen}, {Julbe}, {Karampelas},
  {Kochoska}, {Kohley}, {Kolenberg}, {Kontizas}, {Koposov}, {Kordopatis},
  {Koubsky}, {Kowalczyk}, {Krone-Martins}, {Kudryashova}, {Kull}, {Bachchan},
  {Lacoste-Seris}, {Lanza}, {Lavigne}, {Le Poncin-Lafitte}, {Lebreton},
  {Lebzelter}, {Leccia}, {Leclerc}, {Lecoeur-Taibi}, {Lemaitre}, {Lenhardt},
  {Leroux}, {Liao}, {Licata}, {Lindstr{\o}m}, {Lister}, {Livanou}, {Lobel},
  {L{\"o}ffler}, {L{\'o}pez}, {Lopez-Lozano}, {Lorenz}, {Loureiro},
  {MacDonald}, {Magalh{\~a}es Fernandes}, {Managau}, {Mann}, {Mantelet},
  {Marchal}, {Marchant}, {Marconi}, {Marie}, {Marinoni}, {Marrese},
  {Marschalk{\'o}}, {Marshall}, {Mart{\'\i}n-Fleitas}, {Martino}, {Mary},
  {Matijevi{\v{c}}}, {Mazeh}, {McMillan}, {Messina}, {Mestre}, {Michalik},
  {Millar}, {Miranda}, {Molina}, {Molinaro}, {Molinaro}, {Moln{\'a}r},
  {Moniez}, {Montegriffo}, {Monteiro}, {Mor}, {Mora}, {Morbidelli}, {Morel},
  {Morgenthaler}, {Morley}, {Morris}, {Mulone}, {Muraveva}, {Musella},
  {Narbonne}, {Nelemans}, {Nicastro}, {Noval}, {Ord{\'e}novic},
  {Ordieres-Mer{\'e}}, {Osborne}, {Pagani}, {Pagano}, {Pailler}, {Palacin},
  {Palaversa}, {Parsons}, {Paulsen}, {Pecoraro}, {Pedrosa}, {Pentik{\"a}inen},
  {Pereira}, {Pichon}, {Piersimoni}, {Pineau}, {Plachy}, {Plum}, {Poujoulet},
  {Pr{\v{s}}a}, {Pulone}, {Ragaini}, {Rago}, {Rambaux}, {Ramos-Lerate},
  {Ranalli}, {Rauw}, {Read}, {Regibo}, {Renk}, {Reyl{\'e}}, {Ribeiro},
  {Rimoldini}, {Ripepi}, {Riva}, {Rixon}, {Roelens}, {Romero-G{\'o}mez},
  {Rowell}, {Royer}, {Rudolph}, {Ruiz-Dern}, {Sadowski}, {Sagrist{\`a}
  Sell{\'e}s}, {Sahlmann}, {Salgado}, {Salguero}, {Sarasso}, {Savietto},
  {Schnorhk}, {Schultheis}, {Sciacca}, {Segol}, {Segovia}, {Segransan},
  {Serpell}, {Shih}, {Smareglia}, {Smart}, {Smith}, {Solano}, {Solitro},
  {Sordo}, {Soria Nieto}, {Souchay}, {Spagna}, {Spoto}, {Stampa}, {Steele},
  {Steidelm{\"u}ller}, {Stephenson}, {Stoev}, {Suess}, {S{\"u}veges}, {Surdej},
  {Szabados}, {Szegedi-Elek}, {Tapiador}, {Taris}, {Tauran}, {Taylor},
  {Teixeira}, {Terrett}, {Tingley}, {Trager}, {Turon}, {Ulla}, {Utrilla},
  {Valentini}, {van Elteren}, {Van Hemelryck}, {van Leeuwen}, {Varadi},
  {Vecchiato}, {Veljanoski}, {Via}, {Vicente}, {Vogt}, {Voss}, {Votruba},
  {Voutsinas}, {Walmsley}, {Weiler}, {Weingrill}, {Werner}, {Wevers},
  {Whitehead}, {Wyrzykowski}, {Yoldas}, {{\v{Z}}erjal}, {Zucker}, {Zurbach},
  {Zwitter}, {Alecu}, {Allen}, {Allende Prieto}, {Amorim},
  {Anglada-Escud{\'e}}, {Arsenijevic}, {Azaz}, {Balm}, {Beck}, {Bernstein},
  {Bigot}, {Bijaoui}, {Blasco}, {Bonfigli}, {Bono}, {Boudreault}, {Bressan},
  {Brown}, {Brunet}, {Bunclark}, {Buonanno}, {Butkevich}, {Carret}, {Carrion},
  {Chemin}, {Ch{\'e}reau}, {Corcione}, {Darmigny}, {de Boer}, {de Teodoro}, {de
  Zeeuw}, {Delle Luche}, {Domingues}, {Dubath}, {Fodor}, {Fr{\'e}zouls},
  {Fries}, {Fustes}, {Fyfe}, {Gallardo}, {Gallegos}, {Gardiol}, {Gebran},
  {Gomboc}, {G{\'o}mez}, {Grux}, {Gueguen}, {Heyrovsky}, {Hoar}, {Iannicola},
  {Isasi Parache}, {Janotto}, {Joliet}, {Jonckheere}, {Keil}, {Kim},
  {Klagyivik}, {Klar}, {Knude}, {Kochukhov}, {Kolka}, {Kos}, {Kutka}, {Lainey},
  {LeBouquin}, {Liu}, {Loreggia}, {Makarov}, {Marseille}, {Martayan},
  {Martinez-Rubi}, {Massart}, {Meynadier}, {Mignot}, {Munari}, {Nguyen},
  {Nordlander}, {Ocvirk}, {O'Flaherty}, {Olias Sanz}, {Ortiz}, {Osorio},
  {Oszkiewicz}, {Ouzounis}, {Palmer}, {Park}, {Pasquato}, {Peltzer}, {Peralta},
  {P{\'e}turaud}, {Pieniluoma}, {Pigozzi}, {Poels}, {Prat}, {Prod'homme},
  {Raison}, {Rebordao}, {Risquez}, {Rocca-Volmerange}, {Rosen}, {Ruiz-Fuertes},
  {Russo}, {Sembay}, {Serraller Vizcaino}, {Short}, {Siebert}, {Silva},
  {Sinachopoulos}, {Slezak}, {Soffel}, {Sosnowska}, {Strai{\v{z}}ys}, {ter
  Linden}, {Terrell}, {Theil}, {Tiede}, {Troisi}, {Tsalmantza}, {Tur},
  {Vaccari}, {Vachier}, {Valles}, {Van Hamme}, {Veltz}, {Virtanen}, {Wallut},
  {Wichmann}, {Wilkinson}, {Ziaeepour}, \& {Zschocke}}]{Prusti16}
{Gaia Collaboration}, {Prusti}, T., {de Bruijne}, J.~H.~J., {et~al.} 2016,
  \aap, 595, A1

\bibitem[{{Gaia Collaboration} {et~al.}(2022){Gaia Collaboration}, {Vallenari},
  {Brown}, {Prusti}, {de Bruijne}, {Arenou}, {Babusiaux}, {Biermann},
  {Creevey}, {Ducourant}, {Evans}, {Eyer}, {Guerra}, {Hutton}, {Jordi},
  {Klioner}, {Lammers}, {Lindegren}, {Luri}, {Mignard}, {Panem}, {Pourbaix},
  {Randich}, {Sartoretti}, {Soubiran}, {Tanga}, {Walton}, {Bailer-Jones},
  {Bastian}, {Drimmel}, {Jansen}, {Katz}, {Lattanzi}, {van Leeuwen}, {Bakker},
  {Cacciari}, {Casta{\~n}eda}, {De Angeli}, {Fabricius}, {Fouesneau},
  {Fr{\'e}mat}, {Galluccio}, {Guerrier}, {Heiter}, {Masana}, {Messineo},
  {Mowlavi}, {Nicolas}, {Nienartowicz}, {Pailler}, {Panuzzo}, {Riclet}, {Roux},
  {Seabroke}, {Sordo{\o}rcit}, {Th{\'e}venin}, {Gracia-Abril}, {Portell},
  {Teyssier}, {Altmann}, {Andrae}, {Audard}, {Bellas-Velidis}, {Benson},
  {Berthier}, {Blomme}, {Burgess}, {Busonero}, {Busso}, {C{\'a}novas}, {Carry},
  {Cellino}, {Cheek}, {Clementini}, {Damerdji}, {Davidson}, {de Teodoro},
  {Nu{\~n}ez Campos}, {Delchambre}, {Dell'Oro}, {Esquej},
  {Fern{\'a}ndez-Hern{\'a}ndez}, {Fraile}, {Garabato}, {Garc{\'\i}a-Lario},
  {Gosset}, {Haigron}, {Halbwachs}, {Hambly}, {Harrison}, {Hern{\'a}ndez},
  {Hestroffer}, {Hodgkin}, {Holl}, {Jan{\ss}en}, {Jevardat de Fombelle},
  {Jordan}, {Krone-Martins}, {Lanzafame}, {L{\"o}ffler}, {Marchal}, {Marrese},
  {Moitinho}, {Muinonen}, {Osborne}, {Pancino}, {Pauwels}, {Recio-Blanco},
  {Reyl{\'e}}, {Riello}, {Rimoldini}, {Roegiers}, {Rybizki}, {Sarro}, {Siopis},
  {Smith}, {Sozzetti}, {Utrilla}, {van Leeuwen}, {Abbas}, {{\'A}brah{\'a}m},
  {Abreu Aramburu}, {Aerts}, {Aguado}, {Ajaj}, {Aldea-Montero}, {Altavilla},
  {{\'A}lvarez}, {Alves}, {Anders}, {Anderson}, {Anglada Varela}, {Antoja},
  {Baines}, {Baker}, {Balaguer-N{\'u}{\~n}ez}, {Balbinot}, {Balog}, {Barache},
  {Barbato}, {Barros}, {Barstow}, {Bartolom{\'e}}, {Bassilana}, {Bauchet},
  {Becciani}, {Bellazzini}, {Berihuete}, {Bernet}, {Bertone}, {Bianchi},
  {Binnenfeld}, {Blanco-Cuaresma}, {Blazere}, {Boch}, {Bombrun}, {Bossini},
  {Bouquillon}, {Bragaglia}, {Bramante}, {Breedt}, {Bressan}, {Brouillet},
  {Brugaletta}, {Bucciarelli}, {Burlacu}, {Butkevich}, {Buzzi}, {Caffau},
  {Cancelliere}, {Cantat-Gaudin}, {Carballo}, {Carlucci}, {Carnerero},
  {Carrasco}, {Casamiquela}, {Castellani}, {Castro-Ginard}, {Chaoul},
  {Charlot}, {Chemin}, {Chiaramida}, {Chiavassa}, {Chornay}, {Comoretto},
  {Contursi}, {Cooper}, {Cornez}, {Cowell}, {Crifo}, {Cropper}, {Crosta},
  {Crowley}, {Dafonte}, {Dapergolas}, {David}, {David}, {de Laverny}, {De
  Luise}, {De March}, {De Ridder}, {de Souza}, {de Torres}, {del Peloso}, {del
  Pozo}, {Delbo}, {Delgado}, {Delisle}, {Demouchy}, {Dharmawardena}, {Di
  Matteo}, {Diakite}, {Diener}, {Distefano}, {Dolding}, {Edvardsson}, {Enke},
  {Fabre}, {Fabrizio}, {Faigler}, {Fedorets}, {Fernique}, {Fienga}, {Figueras},
  {Fournier}, {Fouron}, {Fragkoudi}, {Gai}, {Garcia-Gutierrez},
  {Garcia-Reinaldos}, {Garc{\'\i}a-Torres}, {Garofalo}, {Gavel}, {Gavras},
  {Gerlach}, {Geyer}, {Giacobbe}, {Gilmore}, {Girona}, {Giuffrida}, {Gomel},
  {Gomez}, {Gonz{\'a}lez-N{\'u}{\~n}ez}, {Gonz{\'a}lez-Santamar{\'\i}a},
  {Gonz{\'a}lez-Vidal}, {Granvik}, {Guillout}, {Guiraud},
  {Guti{\'e}rrez-S{\'a}nchez}, {Guy}, {Hatzidimitriou}, {Hauser}, {Haywood},
  {Helmer}, {Helmi}, {Sarmiento}, {Hidalgo}, {Hilger}, {H{\l}adczuk}, {Hobbs},
  {Holland}, {Huckle}, {Jardine}, {Jasniewicz}, {Jean-Antoine Piccolo},
  {Jim{\'e}nez-Arranz}, {Jorissen}, {Juaristi Campillo}, {Julbe}, {Karbevska},
  {Kervella}, {Khanna}, {Kontizas}, {Kordopatis}, {Korn}, {K{\'o}sp{\'a}l},
  {Kostrzewa-Rutkowska}, {Kruszy{\'n}ska}, {Kun}, {Laizeau}, {Lambert},
  {Lanza}, {Lasne}, {Le Campion}, {Lebreton}, {Lebzelter}, {Leccia}, {Leclerc},
  {Lecoeur-Taibi}, {Liao}, {Licata}, {Lindstr{\o}m}, {Lister}, {Livanou},
  {Lobel}, {Lorca}, {Loup}, {Madrero Pardo}, {Magdaleno Romeo}, {Managau},
  {Mann}, {Manteiga}, {Marchant}, {Marconi}, {Marcos}, {Marcos Santos},
  {Mar{\'\i}n Pina}, {Marinoni}, {Marocco}, {Marshall}, {Polo},
  {Mart{\'\i}n-Fleitas}, {Marton}, {Mary}, {Masip}, {Massari},
  {Mastrobuono-Battisti}, {Mazeh}, {McMillan}, {Messina}, {Michalik}, {Millar},
  {Mints}, {Molina}, {Molinaro}, {Moln{\'a}r}, {Monari}, {Mongui{\'o}},
  {Montegriffo}, {Montero}, {Mor}, {Mora}, {Morbidelli}, {Morel}, {Morris},
  {Muraveva}, {Murphy}, {Musella}, {Nagy}, {Noval}, {Oca{\~n}a}, {Ogden},
  {Ordenovic}, {Osinde}, {Pagani}, {Pagano}, {Palaversa}, {Palicio},
  {Pallas-Quintela}, {Panahi}, {Payne-Wardenaar}, {Pe{\~n}alosa Esteller},
  {Penttil{\"a}}, {Pichon}, {Piersimoni}, {Pineau}, {Plachy}, {Plum}, {Poggio},
  {Pr{\v{s}}a}, {Pulone}, {Racero}, {Ragaini}, {Rainer}, {Raiteri}, {Rambaux},
  {Ramos}, {Ramos-Lerate}, {Re Fiorentin}, {Regibo}, {Richards}, {Rios Diaz},
  {Ripepi}, {Riva}, {Rix}, {Rixon}, {Robichon}, {Robin}, {Robin}, {Roelens},
  {Rogues}, {Rohrbasser}, {Romero-G{\'o}mez}, {Rowell}, {Royer}, {Ruz Mieres},
  {Rybicki}, {Sadowski}, {S{\'a}ez N{\'u}{\~n}ez}, {Sagrist{\`a} Sell{\'e}s},
  {Sahlmann}, {Salguero}, {Samaras}, {Sanchez Gimenez}, {Sanna},
  {Santove{\~n}a}, {Sarasso}, {Schultheis}, {Sciacca}, {Segol}, {Segovia},
  {S{\'e}gransan}, {Semeux}, {Shahaf}, {Siddiqui}, {Siebert}, {Siltala},
  {Silvelo}, {Slezak}, {Slezak}, {Smart}, {Snaith}, {Solano}, {Solitro},
  {Souami}, {Souchay}, {Spagna}, {Spina}, {Spoto}, {Steele},
  {Steidelm{\"u}ller}, {Stephenson}, {S{\"u}veges}, {Surdej}, {Szabados},
  {Szegedi-Elek}, {Taris}, {Taylo}, {Teixeira}, {Tolomei}, {Tonello}, {Torra},
  {Torra}, {Torralba Elipe}, {Trabucchi}, {Tsounis}, {Turon}, {Ulla}, {Unger},
  {Vaillant}, {van Dillen}, {van Reeven}, {Vanel}, {Vecchiato}, {Viala},
  {Vicente}, {Voutsinas}, {Weiler}, {Wevers}, {Wyrzykowski}, {Yoldas}, {Yvard},
  {Zhao}, {Zorec}, {Zucker}, \& {Zwitter}}]{Vallenari22}
{Gaia Collaboration}, {Vallenari}, A., {Brown}, A.~G.~A., {et~al.} 2022, arXiv
  e-prints, arXiv:2208.00211

\bibitem[{{George} {et~al.}(2016){George}, {Gr{\"a}ff}, {Feuchtgruber},
  {Hartl}, {Eisenhauer}, {Buron}, {Davies}, {Genzel}, {Huber}, {Rau},
  {Plattner}, {Wiezorrek}, {Weisz}, {Amico}, {Glindemann}, {Hau}, \&
  {Kuntschner}}]{George16}
{George}, E.~M., {Gr{\"a}ff}, D., {Feuchtgruber}, H., {et~al.} 2016, in Society
  of Photo-Optical Instrumentation Engineers (SPIE) Conference Series, Vol.
  9908, Ground-based and Airborne Instrumentation for Astronomy VI, ed. C.~J.
  {Evans}, L.~{Simard}, \& H.~{Takami}, 99080G

\bibitem[{{Girardi} {et~al.}(2005){Girardi}, {Groenewegen}, {Hatziminaoglou},
  \& {da Costa}}]{Girardi05}
{Girardi}, L., {Groenewegen}, M.~A.~T., {Hatziminaoglou}, E., \& {da Costa}, L.
  2005, \aap, 436, 895

\bibitem[{{Glikman}(2023)}]{Glikman23}
{Glikman}, E. e.~a. 2023, in prep.

\bibitem[{{Green} {et~al.}(2011){Green}, {Myers}, {Barkhouse}, {Aldcroft},
  {Trichas}, {Richards}, {Ruiz}, \& {Hopkins}}]{Green11}
{Green}, P.~J., {Myers}, A.~D., {Barkhouse}, W.~A., {et~al.} 2011, \apj, 743,
  81

\bibitem[{{Gregg} {et~al.}(2002){Gregg}, {Becker}, {White}, {Richards},
  {Chaffee}, \& {Fan}}]{Gregg02}
{Gregg}, M.~D., {Becker}, R.~H., {White}, R.~L., {et~al.} 2002, \apjl, 573, L85

\bibitem[{{Hennawi} {et~al.}(2010){Hennawi}, {Myers}, {Shen}, {Strauss},
  {Djorgovski}, {Fan}, {Glikman}, {Mahabal}, {Martin}, {Richards}, {Schneider},
  \& {Shankar}}]{Hennawi10}
{Hennawi}, J.~F., {Myers}, A.~D., {Shen}, Y., {et~al.} 2010, \apj, 719, 1672

\bibitem[{{Hennawi} {et~al.}(2006){Hennawi}, {Strauss}, {Oguri}, {Inada},
  {Richards}, {Pindor}, {Schneider}, {Becker}, {Gregg}, {Hall}, {Johnston},
  {Fan}, {Burles}, {Schlegel}, {Gunn}, {Lupton}, {Bahcall}, {Brunner}, \&
  {Brinkmann}}]{Hennawi06}
{Hennawi}, J.~F., {Strauss}, M.~A., {Oguri}, M., {et~al.} 2006, \aj, 131, 1

\bibitem[{{Husemann} {et~al.}(2018){Husemann}, {Worseck}, {Arrigoni Battaia},
  \& {Shanks}}]{Husemann18}
{Husemann}, B., {Worseck}, G., {Arrigoni Battaia}, F., \& {Shanks}, T. 2018,
  \aap, 610, L7

\bibitem[{{Hwang} {et~al.}(2020){Hwang}, {Shen}, {Zakamska}, \&
  {Liu}}]{Hwang20}
{Hwang}, H.-C., {Shen}, Y., {Zakamska}, N., \& {Liu}, X. 2020, \apj, 888, 73

\bibitem[{{Inada} {et~al.}(2008){Inada}, {Oguri}, {Becker}, {Shin}, {Richards},
  {Hennawi}, {White}, {Pindor}, {Strauss}, {Kochanek}, {Johnston}, {Gregg},
  {Kayo}, {Eisenstein}, {Hall}, {Castander}, {Clocchiatti}, {Anderson},
  {Schneider}, {York}, {Lupton}, {Chiu}, {Kawano}, {Scranton}, {Frieman},
  {Keeton}, {Morokuma}, {Rix}, {Turner}, {Burles}, {Brunner}, {Sheldon},
  {Bahcall}, \& {Masataka}}]{Inada08}
{Inada}, N., {Oguri}, M., {Becker}, R.~H., {et~al.} 2008, \aj, 135, 496

\bibitem[{{Inada} {et~al.}(2012){Inada}, {Oguri}, {Shin}, {Kayo}, {Strauss},
  {Morokuma}, {Rusu}, {Fukugita}, {Kochanek}, {Richards}, {Schneider}, {York},
  {Bahcall}, {Frieman}, {Hall}, \& {White}}]{Inada12}
{Inada}, N., {Oguri}, M., {Shin}, M.-S., {et~al.} 2012, \aj, 143, 119

\bibitem[{{Junkkarinen} {et~al.}(2001){Junkkarinen}, {Shields}, {Beaver},
  {Burbidge}, {Cohen}, {Hamann}, \& {Lyons}}]{Junkkarinen01}
{Junkkarinen}, V., {Shields}, G.~A., {Beaver}, E.~A., {et~al.} 2001, \apjl,
  549, L155

\bibitem[{{Lemon} {et~al.}(2020){Lemon}, {Auger}, {McMahon}, {Anguita},
  {Apostolovski}, {Chen}, {Fassnacht}, {Melo}, {Motta}, {Shajib}, {Treu},
  {Agnello}, {Buckley-Geer}, {Schechter}, {Birrer}, {Collett}, {Courbin},
  {Rusu}, {Abbott}, {Allam}, {Annis}, {Avila}, {Bertin}, {Brooks}, {Burke},
  {Carnero Rosell}, {Carrasco Kind}, {Carretero}, {Costanzi}, {da Costa}, {De
  Vicente}, {Desai}, {Eifler}, {Flaugher}, {Frieman}, {Garc{\'\i}a-Bellido},
  {Gaztanaga}, {Gerdes}, {Gruen}, {Gruendl}, {Gschwend}, {Gutierrez},
  {Honscheid}, {James}, {Kim}, {Krause}, {Kuehn}, {Kuropatkin}, {Lahav},
  {Lima}, {Lin}, {Maia}, {March}, {Marshall}, {Menanteau}, {Miquel}, {Palmese},
  {Paz-Chinch{\'o}n}, {Plazas}, {Roodman}, {Sanchez}, {Schubnell}, {Serrano},
  {Smith}, {Soares-Santos}, {Suchyta}, {Tarle}, \& {Walker}}]{Lemon20}
{Lemon}, C., {Auger}, M.~W., {McMahon}, R., {et~al.} 2020, \mnras, 494, 3491

\bibitem[{{Lemon} {et~al.}(2022){Lemon}, {Millon}, {Sluse}, {Courbin}, {Auger},
  {Chan}, {Paic}, \& {Agnello}}]{Lemon22}
{Lemon}, C., {Millon}, M., {Sluse}, D., {et~al.} 2022, \aap, 657, A113

\bibitem[{{Lemon} {et~al.}(2019){Lemon}, {Auger}, \& {McMahon}}]{Lemon19}
{Lemon}, C.~A., {Auger}, M.~W., \& {McMahon}, R.~G. 2019, \mnras, 483, 4242

\bibitem[{{Lemon} {et~al.}(2018){Lemon}, {Auger}, {McMahon}, \&
  {Ostrovski}}]{Lemon18}
{Lemon}, C.~A., {Auger}, M.~W., {McMahon}, R.~G., \& {Ostrovski}, F. 2018,
  \mnras, 479, 5060

\bibitem[{{Li} {et~al.}(2023){Li}, {Liu}, {Shen}, {Oguri}, {Gross}, {Zakamska},
  {Chen}, \& {Hwang}}]{Li23}
{Li}, J., {Liu}, X., {Shen}, Y., {et~al.} 2023, arXiv e-prints,
  arXiv:2306.12502

\bibitem[{{Lockhart} {et~al.}(2019){Lockhart}, {Do}, {Larkin}, {Boehle},
  {Campbell}, {Chappell}, {Chu}, {Ciurlo}, {Cosens}, {Fitzgerald}, {Ghez},
  {Lu}, {Lyke}, {Mieda}, {Rudy}, {Vayner}, {Walth}, \& {Wright}}]{Lockhart19}
{Lockhart}, K.~E., {Do}, T., {Larkin}, J.~E., {et~al.} 2019, \aj, 157, 75

\bibitem[{{Lusso} {et~al.}(2018){Lusso}, {Fumagalli}, {Rafelski}, {Neeleman},
  {Prochaska}, {Hennawi}, {O'Meara}, \& {Theuns}}]{Lusso18}
{Lusso}, E., {Fumagalli}, M., {Rafelski}, M., {et~al.} 2018, \apj, 860, 41

\bibitem[{{Lyke} {et~al.}(2020){Lyke}, {Higley}, {McLane}, {Schurhammer},
  {Myers}, {Ross}, {Dawson}, {Chabanier}, {Martini}, {Busca}, {Mas des
  Bourboux}, {Salvato}, {Streblyanska}, {Zarrouk}, {Burtin}, {Anderson},
  {Bautista}, {Bizyaev}, {Brandt}, {Brinkmann}, {Brownstein}, {Comparat},
  {Green}, {de la Macorra}, {Mu{\~n}oz Guti{\'e}rrez}, {Hou}, {Newman},
  {Palanque-Delabrouille}, {P{\^a}ris}, {Percival}, {Petitjean}, {Rich},
  {Rossi}, {Schneider}, {Smith}, {Vivek}, \& {Weaver}}]{Lyke20}
{Lyke}, B.~W., {Higley}, A.~N., {McLane}, J.~N., {et~al.} 2020, \apjs, 250, 8

\bibitem[{{Mandel} {et~al.}(2007){Mandel}, {Seifert}, {Lenzen}, {Hofmann},
  {J{\"u}tte}, {Weiser}, {Appenzeller}, {Bomans}, {Buschkamp}, {Dettmar},
  {Feiz}, {Gemperlein}, {Germeroth}, {Grimm}, {Heidt}, {Knierim}, {Laun},
  {Lehmitz}, {Luks}, {Mall}, {Polsterer}, {Schimmelmann}, {Weisz}, \&
  {Quirrenbach}}]{Mandel07}
{Mandel}, H., {Seifert}, W., {Lenzen}, R., {et~al.} 2007, Astronomische
  Nachrichten, 328, 626

\bibitem[{{Mannucci} {et~al.}(2022){Mannucci}, {Pancino}, {Belfiore}, {Cicone},
  {Ciurlo}, {Cresci}, {Lusso}, {Marasco}, {Marconi}, {Nardini}, {Pinna},
  {Severgnini}, {Saracco}, {Tozzi}, \& {Yeh}}]{Mannucci22}
{Mannucci}, F., {Pancino}, E., {Belfiore}, F., {et~al.} 2022, Nature Astronomy,
  6, 1185

\bibitem[{{More} {et~al.}(2016){More}, {Oguri}, {Kayo}, {Zinn}, {Strauss},
  {Santiago}, {Mosquera}, {Inada}, {Kochanek}, {Rusu}, {Brownstein}, {da
  Costa}, {Kneib}, {Maia}, {Quimby}, {Schneider}, {Streblyanska}, \&
  {York}}]{More16}
{More}, A., {Oguri}, M., {Kayo}, I., {et~al.} 2016, \mnras, 456, 1595

\bibitem[{{Morgan} {et~al.}(2000){Morgan}, {Burley}, {Costa}, {Maza},
  {Persson}, {Ruiz}, {Schechter}, {Thompson}, \& {Winn}}]{Morgan00}
{Morgan}, N.~D., {Burley}, G., {Costa}, E., {et~al.} 2000, Astronomical
  Journal, 119, 1083

\bibitem[{{Pindor} {et~al.}(2006){Pindor}, {Eisenstein}, {Gregg}, {Becker},
  {Inada}, {Oguri}, {Hall}, {Johnston}, {Richards}, {Schneider}, {Turner},
  {Brasi}, {Hinz}, {Kenworthy}, {Miller}, {Barentine}, {Brewington},
  {Brinkmann}, {Harvanek}, {Kleinman}, {Krzesinski}, {Long}, {Neilsen},
  {Newman}, {Nitta}, {Snedden}, \& {York}}]{Pindor06}
{Pindor}, B., {Eisenstein}, D.~J., {Gregg}, M.~D., {et~al.} 2006, \aj, 131, 41

\bibitem[{{Pinna} {et~al.}(2016){Pinna}, {Esposito}, {Hinz}, {Agapito},
  {Bonaglia}, {Puglisi}, {Xompero}, {Riccardi}, {Briguglio}, {Arcidiacono},
  {Carbonaro}, {Fini}, {Montoya}, \& {Durney}}]{Pinna16}
{Pinna}, E., {Esposito}, S., {Hinz}, P., {et~al.} 2016, in Society of
  Photo-Optical Instrumentation Engineers (SPIE) Conference Series, Vol. 9909,
  Adaptive Optics Systems V, ed. E.~{Marchetti}, L.~M. {Close}, \& J.-P.
  {V{\'e}ran}, 99093V

\bibitem[{{Pinna} {et~al.}(2021){Pinna}, {Rossi}, {Puglisi}, {Agapito},
  {Bonaglia}, {Plantet}, {Mazzoni}, {Briguglio}, {Carbonaro}, {Xompero},
  {Grani}, {Riccardi}, {Esposito}, {Hinz}, {Vaz}, {Ertel}, {Montoya}, {Durney},
  {Christou}, {Miller}, {Taylor}, {Cavallaro}, \& {Lefebvre}}]{Pinna21}
{Pinna}, E., {Rossi}, F., {Puglisi}, A., {et~al.} 2021, arXiv e-prints,
  arXiv:2101.07091

\bibitem[{{Piotto} {et~al.}(2015){Piotto}, {Milone}, {Bedin}, {Anderson},
  {King}, {Marino}, {Nardiello}, {Aparicio}, {Barbuy}, {Bellini}, {Brown},
  {Cassisi}, {Cool}, {Cunial}, {Dalessandro}, {D'Antona}, {Ferraro}, {Hidalgo},
  {Lanzoni}, {Monelli}, {Ortolani}, {Renzini}, {Salaris}, {Sarajedini}, {van
  der Marel}, {Vesperini}, \& {Zoccali}}]{Piotto15}
{Piotto}, G., {Milone}, A.~P., {Bedin}, L.~R., {et~al.} 2015, \aj, 149, 91

\bibitem[{{Rosas-Guevara} {et~al.}(2019){Rosas-Guevara}, {Bower}, {McAlpine},
  {Bonoli}, \& {Tissera}}]{Rosas-Guevara2019}
{Rosas-Guevara}, Y.~M., {Bower}, R.~G., {McAlpine}, S., {Bonoli}, S., \&
  {Tissera}, P.~B. 2019, \mnras, 483, 2712

\bibitem[{{Schechter} {et~al.}(2017){Schechter}, {Morgan}, {Chehade},
  {Metcalfe}, {Shanks}, \& {McDonald}}]{Scheckter17}
{Schechter}, P.~L., {Morgan}, N.~D., {Chehade}, B., {et~al.} 2017, \aj, 153,
  219

\bibitem[{{Scialpi} {et~al.}(2023){Scialpi}, {Mannucci}, \& et~al.}]{Scialpi23}
{Scialpi}, M., {Mannucci}, F., \& et~al. 2023, submitted

\bibitem[{{Shen} {et~al.}(2023){Shen}, {Hwang}, {Oguri}, {Chen}, {Di Matteo},
  {Ni}, {Bird}, {Zakamska}, {Liu}, {Chen}, \& {Kratter}}]{Shen23}
{Shen}, Y., {Hwang}, H.-C., {Oguri}, M., {et~al.} 2023, \apj, 943, 38

\bibitem[{{Shen} {et~al.}(2019){Shen}, {Hwang}, {Zakamska}, \& {Liu}}]{Shen19}
{Shen}, Y., {Hwang}, H.-C., {Zakamska}, N., \& {Liu}, X. 2019, \apjl, 885, L4

\bibitem[{{Silverman} {et~al.}(2020){Silverman}, {Tang}, {Lee}, {Hartwig},
  {Goulding}, {Strauss}, {Schramm}, {Ding}, {Riffel}, {Fujimoto}, {Hikage},
  {Imanishi}, {Iwasawa}, {Jahnke}, {Kayo}, {Kashikawa}, {Kawaguchi}, {Kohno},
  {Luo}, {Matsuoka}, {Matsuda}, {Nagao}, {Oguri}, {Ono}, {Onoue}, {Ouchi},
  {Shimasaku}, {Suh}, {Suzuki}, {Taniguchi}, {Toba}, {Ueda}, \&
  {Yasuda}}]{Silverman20}
{Silverman}, J.~D., {Tang}, S., {Lee}, K.-G., {et~al.} 2020, \apj, 899, 154

\bibitem[{{Simioni} {et~al.}(2018){Simioni}, {Bedin}, {Aparicio}, {Piotto},
  {Milone}, {Nardiello}, {Anderson}, {Bellini}, {Brown}, {Cassisi}, {Cunial},
  {Granata}, {Ortolani}, {van der Marel}, \& {Vesperini}}]{Simioni18}
{Simioni}, M., {Bedin}, L.~R., {Aparicio}, A., {et~al.} 2018, \mnras, 476, 271

\bibitem[{{Skrutskie} {et~al.}(2006){Skrutskie}, {Cutri}, {Stiening},
  {Weinberg}, {Schneider}, {Carpenter}, {Beichman}, {Capps}, {Chester},
  {Elias}, {Huchra}, {Liebert}, {Lonsdale}, {Monet}, {Price}, {Seitzer},
  {Jarrett}, {Kirkpatrick}, {Gizis}, {Howard}, {Evans}, {Fowler}, {Fullmer},
  {Hurt}, {Light}, {Kopan}, {Marsh}, {McCallon}, {Tam}, {Van Dyk}, \&
  {Wheelock}}]{Skrutskie06}
{Skrutskie}, M.~F., {Cutri}, R.~M., {Stiening}, R., {et~al.} 2006, \aj, 131,
  1163

\bibitem[{{Steinborn} {et~al.}(2016){Steinborn}, {Dolag}, {Comerford},
  {Hirschmann}, {Remus}, \& {Teklu}}]{Steinborn2016}
{Steinborn}, L.~K., {Dolag}, K., {Comerford}, J.~M., {et~al.} 2016, \mnras,
  458, 1013

\bibitem[{{Tang} {et~al.}(2021){Tang}, {Silverman}, {Ding}, {Li}, {Lee},
  {Strauss}, {Goulding}, {Schramm}, {Kawinwanichakij}, {Xavier Prochaska},
  {Hennawi}, {Imanishi}, {Iwasawa}, {Toba}, {Kayo}, {Oguri}, {Matsuoka},
  {Onoue}, {Jahnke}, {Ichikawa}, {Hartwig}, {Kashikawa}, {Kawaguchi}, {Kohno},
  {Matsuda}, {Nagao}, {Ono}, {Ouchi}, {Shimasaku}, {Suh}, {Suzuki},
  {Taniguchi}, {Ueda}, \& {Yasuda}}]{Tang21}
{Tang}, S., {Silverman}, J.~D., {Ding}, X., {et~al.} 2021, \apj, 922, 83

\bibitem[{{Temple} {et~al.}(2021){Temple}, {Hewett}, \& {Banerji}}]{Temple2021}
{Temple}, M.~J., {Hewett}, P.~C., \& {Banerji}, M. 2021, \mnras, 508, 737

\bibitem[{{Tremmel} {et~al.}(2017){Tremmel}, {Karcher}, {Governato},
  {Volonteri}, {Quinn}, {Pontzen}, {Anderson}, \& {Bellovary}}]{Tremmel17}
{Tremmel}, M., {Karcher}, M., {Governato}, F., {et~al.} 2017, \mnras, 470, 1121

\bibitem[{{Vanden Berk} {et~al.}(2001){Vanden Berk}, {Richards}, {Bauer},
  {Strauss}, {Schneider}, {Heckman}, {York}, {Hall}, {Fan}, {Knapp},
  {Anderson}, {Annis}, {Bahcall}, {Bernardi}, {Briggs}, {Brinkmann}, {Brunner},
  {Burles}, {Carey}, {Castander}, {Connolly}, {Crocker}, {Csabai}, {Doi},
  {Finkbeiner}, {Friedman}, {Frieman}, {Fukugita}, {Gunn}, {Hennessy},
  {Ivezi{\'c}}, {Kent}, {Kunszt}, {Lamb}, {Leger}, {Long}, {Loveday}, {Lupton},
  {Meiksin}, {Merelli}, {Munn}, {Newberg}, {Newcomb}, {Nichol}, {Owen}, {Pier},
  {Pope}, {Rockosi}, {Schlegel}, {Siegmund}, {Smee}, {Snir}, {Stoughton},
  {Stubbs}, {SubbaRao}, {Szalay}, {Szokoly}, {Tremonti}, {Uomoto}, {Waddell},
  {Yanny}, \& {Zheng}}]{Vandenberk01}
{Vanden Berk}, D.~E., {Richards}, G.~T., {Bauer}, A., {et~al.} 2001, \aj, 122,
  549

\bibitem[{{Vanhollebeke} {et~al.}(2009){Vanhollebeke}, {Groenewegen}, \&
  {Girardi}}]{Vanhollebeke09}
{Vanhollebeke}, E., {Groenewegen}, M.~A.~T., \& {Girardi}, L. 2009, \aap, 498,
  95

\bibitem[{{Volonteri} {et~al.}(2022){Volonteri}, {Pfister}, {Beckmann},
  {Dotti}, {Dubois}, {Massonneau}, {Musoke}, \& {Tremmel}}]{Volonteri21}
{Volonteri}, M., {Pfister}, H., {Beckmann}, R., {et~al.} 2022, \mnras, 514, 640

\bibitem[{{Yao} {et~al.}(2019){Yao}, {Wu}, {Ai}, {Yang}, {Yang}, {Dong},
  {Joshi}, {Wang}, {Feng}, {Fu}, {Hou}, {Luo}, {Kong}, {Liu}, {Zhao}, {Zhang},
  {Yuan}, \& {Shen}}]{Yao19}
{Yao}, S., {Wu}, X.-B., {Ai}, Y.~L., {et~al.} 2019, \apjs, 240, 6

\end{thebibliography}



\appendix 

\section{Results of the spectral deconvolution}

\begin{table*}
\caption[]{Best fit parameters of the spectral deconvolution}
\label{tab:deconvolution}
\centering
\begin{tabular}{ccclcccc}
\hline
 Target          &   ~z$_{fit}$      & E$_{(B-V)}$ & G$_{P, fit}$ & G$_{S, fit}$& Star type & V$_{*}$ \\
                &                 &                  &          &        & & [km/s]&  \\ 
\hline
\mr{J0732+3533}  &\mr{3.0529}&\mr{0.0698}& &\\
                 &                &                  &          & & & & \\
\hline
\mr{J0812-0040}  &\mr{1.9147}&\mr{0.0004}&\mr{20.36} &\mr{20.39}& \mr{G0}      & \mr{+286}          & \\                 &                &       &          & & & \\
\hline
\mr{J0812+2007}  &\mr{1.4865}&\mr{0.0358}&\mr{20.16} &\mr{21.46}& \mr{G2} & \mr{+57}\\
                 &                &                  &          & & \\
\hline
\mr{J0927+33512}  &\mr{1.1605}&\mr{0.0000}& &&  &  \\
                 &                &                  &          & & \\
\hline
\mr{J0950+4329}  &\mr{1.7700}&\mr{0.1389}& &&  &  \\
                 &                &                  &          & & \\
\hline
\mr{J1048+4541}  &\mr{1.4432}&\mr{0.0449}& &\\
                 &                &                  &          & & & & \\
                 
\hline
\mr{J1103+2348}   &\mr{1.4384}&\mr{0.1003}& &\\
                 &                &                  &          & & & & \\
                 \hline
\mr{J1318-0136}  &\mr{1.4881}&\mr{0.1498}& &&  &  \\
                 &                &                  &          & & \\
\hline
\mr{J1510+5959}  &\mr{2.0212}&\mr{0.0039}&\mr{20.34} &\mr{20.99}& \mr{G3} & \mr{-210}  \\
                 &                &                  &          & & \\
\hline
\end{tabular}
\tablefoot{Best fit parameters for the deconvolved sources. From left to right: target ID, spectroscopic redshift (z$_{fit}$), and extinction (E$_{(B-V)}$) of the best-fitting QSO template. In case an object is reproduced by and AGN plus a star, we report the  G-band magnitude of the primary and secondary objects (G$_{P, fit})$ and G$_{S, fit}$, respectively), the spectral type of the best-fitting star, and its radial velocity  (V$_{*}$)}.\\
\end{table*}

\begin{figure*}
\centering
\includegraphics[width=18cm]{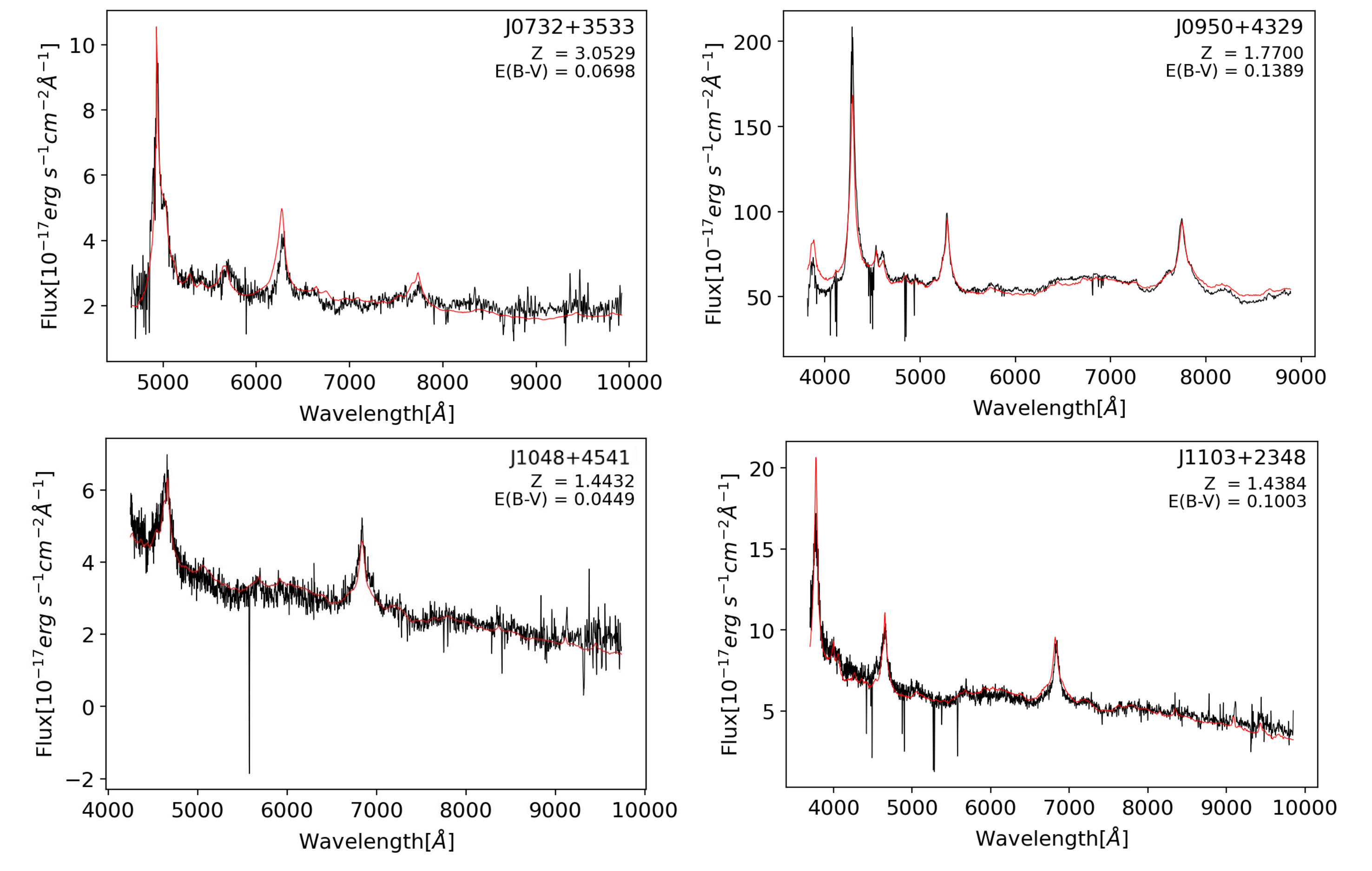}
\caption{
Spectral deconvolution of the objects best reproduced by an AGN spectrum. The observed spectrum is shown in black, the best fit is in red. Each panel reports the target ID, and the best fitting redshift and amount of dust extinction.
}
\label{fig:deconv1}
\end{figure*}

\begin{figure*}
\centering
\includegraphics[width=0.7\textwidth]{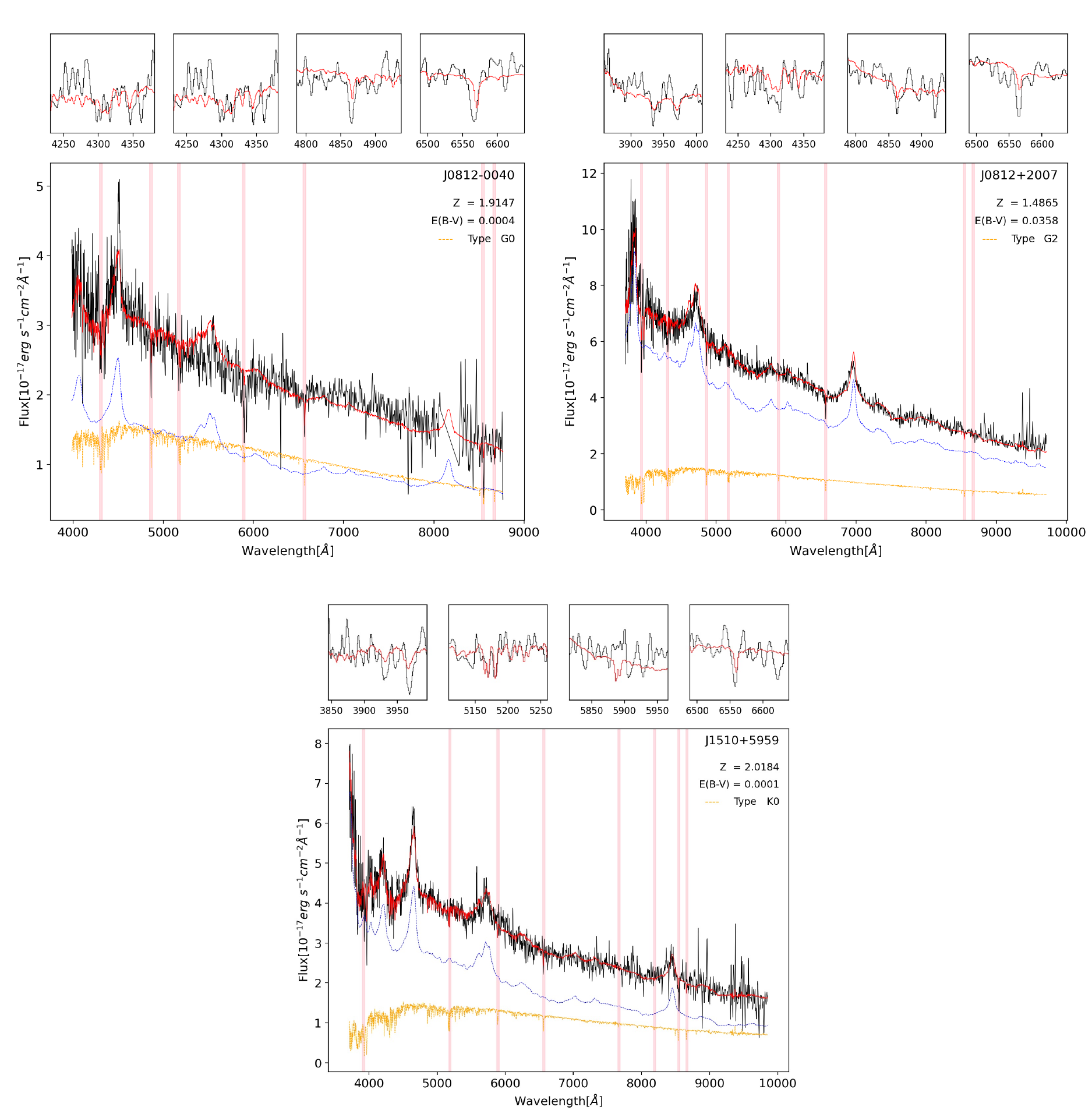}
\caption{
Spectral deconvolution of the systems best reproduced by a combination of an AGN and a star. Observed spectra are shown in black, the best fit in red, the QSO in blue, the stellar component in orange, with red vertical stripes around the main stellar absorption features of the best-fitting stellar type. The four small panels above the main ones are enlargements of the main absorptions features of the stellar companion.
}
\label{fig:deconv2}
\end{figure*}

\begin{figure*}
\centering
\includegraphics[width=18cm]{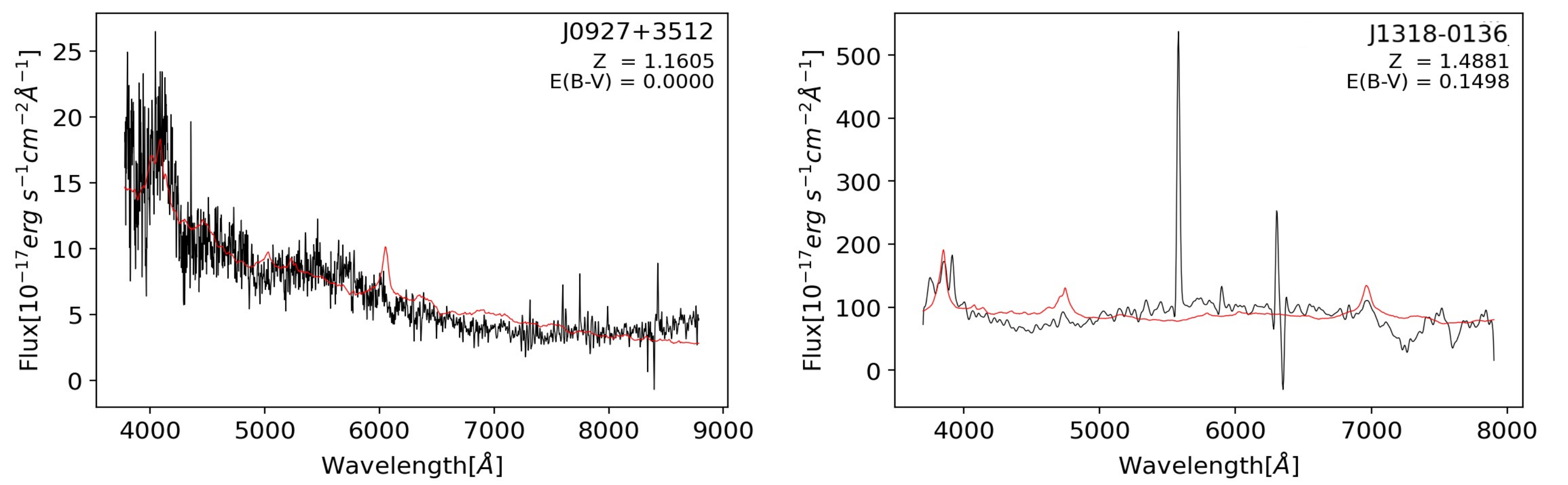}
\caption{
Spectral deconvolution of the unclassified objects, i.e., the systems whose spectra are not well reproduced neither from an AGN, nor from an AGN plus a star.
}
\label{fig:deconv3}
\end{figure*}

\end{document}